\input harvmac
\overfullrule=0mm

%
\def\frac#1#2{\scriptstyle{#1 \over #2}}

\def\ket#1{ | #1 \rangle}
\def\bra#1{ \langle #1 |}
%
%

\def\CV{{\cal V}}		

\def\({ \left( }\def\[{ \left[ }
\def\){ \right) }\def\]{ \right] }
%


\def\IR{\relax{\rm I\kern-.18em R}}
\font\cmss=cmss10 \font\cmsss=cmss10 at 7pt
\def\IZ{\relax\ifmmode\mathchoice
{\hbox{\cmss Z\kern-.4em Z}}{\hbox{\cmss Z\kern-.4em Z}}
{\lower.9pt\hbox{\cmsss Z\kern-.4em Z}}
{\lower1.2pt\hbox{\cmsss Z\kern-.4em Z}}\else{\cmss Z\kern-.4em Z}\fi}
\def\inbar{\,\vrule height1.5ex width.4pt depth0pt}
\def\IB{\relax{\rm I\kern-.18em B}}
\def\IC{\relax\hbox{$\inbar\kern-.3em{\rm C}$}}
\def\ID{\relax{\rm I\kern-.18em D}}
\def\IE{\relax{\rm I\kern-.18em E}}
\def\IF{\relax{\rm I\kern-.18em F}}
\def\IG{\relax\hbox{$\inbar\kern-.3em{\rm G}$}}
\def\IH{\relax{\rm I\kern-.18em H}}
\def\II{\relax{\rm I\kern-.18em I}}
\def\IK{\relax{\rm I\kern-.18em K}}
\def\IL{\relax{\rm I\kern-.18em L}}
\def\IM{\relax{\rm I\kern-.18em M}}
\def\IN{\relax{\rm I\kern-.18em N}}
\def\IO{\relax\hbox{$\inbar\kern-.3em{\rm O}$}}
\def\IP{\relax{\rm I\kern-.18em P}}
\def\IQ{\relax\hbox{$\inbar\kern-.3em{\rm Q}$}}
\def\IGa{\relax\hbox{${\rm I}\kern-.18em\Gamma$}}
\def\IPi{\relax\hbox{${\rm I}\kern-.18em\Pi$}}
\def\ITh{\relax\hbox{$\inbar\kern-.3em\Theta$}}
\def\IOm{\relax\hbox{$\inbar\kern-3.00pt\Omega$}}



\def\oh{{1\over 2}}\def\un{{\bf 1}}


\def\Ge{\epsilon}
\def\Gth{\theta}
\def\Gl{\lambda}

\def\Gr{\rho}
\def\Gt{\tau}


\def\mod{{\rm mod\,}}
\def\diag{{\rm diag \,}}
 
\def\bra{\langle}\def\ket{\rangle}
\def\tt{t_{\textstyle{.}}}

 \def\Che{Chebishev\ } 

\Title{SPhT 93/057;  hep-th/9306018}
{{\vbox {
\centerline{Graph Rings and  Integrable Perturbations} 
\bigskip
\centerline{of $N=2$ Superconformal Theories} }}}

\bigskip

\centerline{P. Di Francesco,} \bigskip \centerline{F. Lesage} \bigskip
\centerline{and}\bigskip\centerline{J.-B. Zuber}\bigskip

\centerline{ \it Service de Physique Th\'eorique de Saclay
\footnote*{Laboratoire de la Direction des Sciences et
de la Mati\`ere du Commissariat \`a l'Energie Atomique.},}
\centerline{ \it F-91191 Gif sur Yvette Cedex, France}

\vskip .2in

\noindent 
We show that the connection between certain integrable perturbations
of $N=2$ superconformal theories and graphs found by Lerche and Warner
extends to a broader class. These perturbations 
are such that the generators of the perturbed chiral ring may be 
diagonalized in an orthonormal basis. This allows to define a dual 
ring, whose generators are labelled by the ground states of the theory and 
are encoded in a graph or set of graphs, that reproduce the pattern of the
ground states and interpolating solitons. All known perturbations of the 
$ADE$ potentials and some others are shown to satisfy this criterion. 
This suggests a test of integrability.

\Date{6/93\qquad\qquad to be submitted to Nuclear Physics B.}
%

\lref\EV{E. Verlinde, Nucl. Phys. {\bf B300} [FS22] 360-376 (1988). }
\lref\Arn{V.I. Arnold,  and S.M. Gusein-Zade, A.N. Varchenko,
{\it Singularities of differentiable maps}, Birk\"auser, Basel 1985.}
\lref\KZ{Y. Kazama and H. Suzuki, Phys. Lett. {\bf B216} 112-116 (1989); 
Nucl. Phys. {\bf B321} 232-268 (1989). }
\lref\VWLM{E. Martinec, Phys. Lett. {\bf B217} 431-437 (1989); {\it Criticality, 
catastrophes and compactifications}, in 
{\it Physics and mathematics of strings}, V.G. Knizhnik memorial volume, 
L. Brink, D. Friedan and A.M. Polyakov eds., World Scientific 1990 
 \semi C. Vafa and  N.P. Warner, Phys. Lett. {\bf B218} 51-58 (1989).}
\lref\LVW{W. Lerche, C. Vafa, N.P. Warner, Nucl. Phys. {\bf B324} 427-474  
(1989).}
\lref\DVV{R. Dijkgraaf, E. Verlinde and H. Verlinde, Nucl. Phys. 
{\bf B352} 59-86 (1991).}
\lref\NW{D. Nemeschansky and N.P. Warner, Nucl. Phys. {\bf B380} 241-264 (1992).}
\lref\FLMW{P. Fendley, W. Lerche, S.D. Mathur and N.P. Warner,
Nucl. Phys. {\bf B348} 66-88 (1991) .}
\lref\FMVW{P. Fendley, S.D. Mathur, C. Vafa and N.P. Warner, 
Phys. Lett. {\bf 243B} 257-264 (1991). }
\lref\LW{W. Lerche and N.P. Warner, 
in {\it Strings \& Symmetries, 1991}, N. Berkovits, H. Itoyama et al. eds, 
World Scientific 1992.}
\lref\LWnp{W. Lerche and N.P. Warner, 
 Nucl. Phys. {\bf B358} 571-599 (1991).}
\lref\Wa{N. Warner, {\it $N=2$ Supersymmetric Integrable Models and
Topological Field Theories}, to appear in the proceedings of the 1992 
Trieste Summer School, hep-th/9301088. }
\lref\KTS{S. Mahapatra, Phys. Lett. {\bf B264} (1991) 50-56\semi
A. Klemm, S. Theisen und M.G. Schmidt, Int.J.Mod.Phys. {\bf A7}
6215-6244 (1992).}
\lref\Maha{S. Mahapatra, Phys. Lett. {\bf B264} (1991) 50-56.}
\lref\DFZ{P. Di Francesco and J.-B. Zuber, 
in {\it Recent Developments in Conformal Field Theories}, Trieste
Conference, 1989, S. Randjbar-Daemi, E. Sezgin and J.-B. Zuber eds., 
World Scientific 1990
\semi P. Di Francesco, Int.J.Mod.Phys. {\bf A7} 407-500 (1992).}
\lref\Dub{B. Dubrovin, Nucl. Phys. {\bf B 379} 627-689 (1992); 
{\it Differential Geometry of the space of orbits of a Coxeter group},
preprint hep-th/9303152.}
\lref\DFZf{P. Di Francesco and J.-B. Zuber, {\it Fusion potentials I}, 
J.Phys. {\bf A 26}, 1441-1454 (1993). }
\lref\Dij{R. Dijkgraaf, {\it Intersection Theory, Integrable Hierarchies
and Topological Field Theory}, to appear in 
{\it New Symmetry Principles in Quantum Field Theory}, 
proceedings of the Carg\`ese Summer School July 1991, hep-th/9201003. }
\lref\NWZ{D. Nemeshansky, N.P. Warner, and J.-B. Zuber, in preparation.}
\lref\CV{S. Cecotti and C. Vafa, 
{\it On Classification of $N=2$ Supersymmetric Theories}, 
preprint HUTP-92/A064,SISSA-203/92/EP, hep-th/9211097.  }
\lref\CVtat{S. Cecotti and C. Vafa, 
Nucl. Phys. {\bf B367} 359-461 (1991).}
\lref\FIi{P. Fendley and K. Intriligator, Nucl. Phys. {\bf B372} 533-558 
(1992).} 
\lref\FIii{P. Fendley and K. Intriligator, Nucl. Phys. 
{\bf B380} 265-290 (1992).}
\lref\DG{D. Gepner, {\it Foundations of Rational Quantum Field Theory, I}
preprint CALT-68-1825, hep-th/9211100. }
\lref\VP{V. Pasquier, J.Phys. {\bf A20} 5707-5717 (1987)\semi
P. Ginsparg, {\it Some statistical mechanical models and conformal
 field theories}, lectures at the Trieste spring school, April 1989.}
\lref\OEK{A. Ocneanu in 
{\it Operator Algebras and Applications}, vol.2, 119-172, London Math. Soc. 
Lecture Notes Series, Vol. 136, Cambridge Univ. Press, London 1988; 
{\it Quantum symmetry, differential geometry of finite graphs and 
classification of subfactors}, Univ. of Tokyo Seminar Notes 45, notes taken by
Y. Kawahigashi
\semi  Y. Kawahigashi, {\it On flatness of Oneanu's connections on the 
Dynkin diagram and classification of subfactors}, preprint 1990
\semi
D. Evans and Y. Kawahigashi, {\it The $E_7$ commuting squares produce $D_{10}$
as principal graph}, preprint 1992.}
\lref\Bou{M. Bourdeau, E.J. Mlawer, H. Riggs and H.J. Schnitzer, Mod. Phys. Lett. {\bf A7} 698-700 (1992).}
\lref\GSc{D. Gepner and A. Schwimmer, Nucl. Phys. {\bf B380} 147-167 (1992).}
\lref\VeW{H. Verlinde and N.P. Warner, Phys. Lett. {\bf B 269} 96-102 (1991).}
\lref\MW{P. Mathieu and M.A. Walton, Phys. Lett. {\bf B 254} 106-110 (1991).}
\lref\EY{T. Eguchi and S.-K. Yang, Mod. Phys. lett. {\bf A5} 1693 (1990).}
\lref\Wit{E. Witten, {\it On the Landau-Ginzburg description of $N=2$
minimal models}, preprint IASSNS-HEP-93/10, hep-th/9304026.}
\lref\DFY{P. Di Francesco and S. Yankielowicz, {\it Ramond sector characters
and $N=2$ Landau-Ginzburg models}, preprint T93/049, TAUP 2047-93, hep-th/9305037.}
\lref\BI{E. Bannai and T. Ito, {\it Algebraic Combinatorics I: Association
Schemes}, Benjamin-Cummings, 1984.}
\lref\GAN{Ph. Ruelle, E. Thiran and J. Weyers, 
{\it Implications of an arithmetical symmetry of the commutant for 
modular invariants}, preprint hep-th/9212048\semi 
T. Gannon, {\it The classification of Affine $SU(3)$ Modular 
Invariant Partition Functions}, preprint hep-th/9212060.}
\lref\DGe{D. Gepner, Comm. Math. Phys. {\bf 141} 381-411 (1991).}
\lref\Vin{V. Pasquier, Th\`ese d'Etat, Orsay, 1988. }


\newsec{Introduction}

\noindent
$N=2$ superconformal theories have been under active investigation during 
recent years. Of particular interest is to understand their perturbations
by relevant operators, that preserve the $N=2$ supersymmetry and make them
massive field theories, and among them, those that are integrable (\FMVW, 
for a review and a list of references see \Wa).  Integrability seems to imply nice features on the 
pattern of solitons that interpolate between the ground states of the theory. 

Two years ago, Lerche and Warner have studied the perturbations of the  
$N=2$ theories described by an $ADE$ Landau-Ginzburg potential 
\refs{\VWLM,\LVW} and perturbed 
by the least relevant operator \LW. They made the intriguing observation 
that the pattern of minima of the potential in field space then reproduces 
the shape of the corresponding Dynkin diagram, and that the values of the
polynomial representatives of the chiral ring at the minima of the potential
are proportional to the various eigenvectors of the Cartan matrix. In 
the $A$ case, the chiral algebra satisfied by these polynomials is
nothing else than the Verlinde fusion algebra of the corresponding $SU(2)$
theory, whereas in the other $D$ or $E$ cases, its interpretation 
remained more elusive. 

In this paper, we want to point out that this feature extends to a much
broader class of integrable perturbations, that the association between 
the corresponding ring (or algebra)
 and a graph is something that has already been 
encountered some time ago in the slightly different context of relations
between lattice integrable models attached to graphs and conformal field 
theories\DFZ\ and that conversely this connection might suggest a simple 
criterion of integrability. 
In Sec. 2, we review shortly some facts on $N=2$ superconformal 
field theories, their perturbations and the structure of their ``chiral 
ring''. We focus on cases 
where the multiplication matrices $C_i$ in a natural basis of the (perturbed) 
chiral ring are {\it normal}, {\it i.e.} commute with their
adjoint $[C_i,C_i^{\dagger}]=0$. 
This condition is equivalent to the property of $C_i$ to be diagonalizable
in an orthonormal basis\foot{
A simple proof of this fact is provided by
the decomposition of $C$ into its hermitian and antihermitian parts, 
$C=A+iB$, $A=A^{\dagger}$, $B=B^{\dagger}$; the commutation of $C$ and
$C^{\dagger}$ translates into the commutation of $A$ and $B$ that may thus
be diagonalized simultaneously in an orthonormal basis.}.  
More generally, we assume that the 
matrices  may be made normal after a change of basis that
respects the natural gradation of the problem ($U(1)$ charge). This 
we call the {\it normalizability } property. 
We then recall in Sec. 3 some definitions of what we 
call a {\it graph ring}, of its dual ring, and how in many cases
a subring of it enables one to identify a structure of blocks with nice 
modular properties. Applied to the $ADE$ perturbations mentionned above, 
it tells us how to 
group the polynomials into blocks in one-to-one correspondence 
with the blocks of the corresponding $SU(2)$ 
modular invariant.  A survey of the $ADE$ cases (sect. 
4) and of cases related to $SU(N)$ (sect. 5) 
shows that the property of normalizability is quite restrictive and 
allows only sparse solutions. Strangely, all the known integrable cases
of $ADE$ potentials are normalizable, and the dual ring has integral 
coefficients that may thus be encoded in a graph. In so far as the normal 
matrices are some natural generalization of the fusion matrices, 
this connection 
between the property of normalizability and integrability represents an 
extension of a conjecture of Gepner \DG\ 
that there is an underlying conformal 
field theory behind each integrable deformation of $N=2$ theory. 
Moreover, this condition 
of normalizability enables us to identify some plausible 
candidates to integrability. One of them has been checked to  
possess indeed conserved quantities \NWZ. 
We speculate in sect. 6 on the issues raised in this paper.
Three appendices gather some technical material on the form of
the $D$ potential and free energy (App. A), on the exceptional cases
(App. B) and on $SU(3)$ at level 2 (App. C). 


\newsec{Perturbations of $N=2$ Landau-Ginzburg superconformal theories}
\noindent Consider a $N=2$ superconformal theory that admits a 
description by a Landau-Ginzburg superpotential. The latter is a 
quasi-homogeneous polynomial in the superfields, $W_0(x,y,\cdots)$. 
The case of the ``minimal'' $N=2$ theories (with $c<3$) is particularly
striking since this superpotential is  given by one of the well-known
$ADE$ singularities \Arn, thus matching 
the classification of modular invariants \refs{\VWLM,\LVW}. 
This description is 
substantiated by the comparison of the chiral ring of the $N=2$
theory with the local ring of the singularity. The former describes the
non-singular pointwise product of fields that satisfy the constraint
$h=\oh q$, with $h$ their conformal weight and $q$ their $U(1)$ charge, and 
the latter describes the multiplication in the ring of polynomials 
$\mod \partial_x W_{0}, \partial_y W_0, \cdots$. 

The important question of the $N=2$ supersymmetry preserving perturbations
\EY\ may then be investigated using this potential description.

Let $W(x, \dots, \tt)$ be the perturbed potential in terms of the 
flat coordinates $t_i$, (see \Wa\ for a definition and references), 
let $p_l(x,\dots, \tt)=- {\partial W\over \partial t_l}$ 
be the corresponding basis of the (deformed) chiral ring with $p_0=1$. 
The structure constants of the ring
\eqn\IIaa{ p_i p_j =C_{ij}^{\ \ k} p_k\qquad \mod \partial W }
are functions of the coordinates $\tt$ and have been proved to 
satisfy two kinds of constraints, (in addition to the associativity
and commutativity which are obvious from that definition) \DVV
\item{$\bullet$} the metric tensor defined as $\eta_{ij}=C_{ij}^{\ \ 0}$
is independent of the $t$'s; 
\item{$\bullet$} $C_{ij}^{ \ \ k}$ 
satisfy the integrability condition that enable
one to write $C_{ij}^{\ \ k}= 
{\partial^3\over \partial t_i\partial t_j\partial t_k}
F(\tt)$, where $F(\tt)$ is some function, the free energy of the theory. 

\noindent We should mention that 
Dubrovin has undertaken the classification of the solutions to these 
constraints, independently of the existence of a potential and polynomial
representation of the chiral ring \Dub. 
For more on these topics, see \Dij.

For any given perturbation, let us consider the chiral ring. Consider
first for simplicity the case where the potential depends on a single
variable $x$. Let $C_1$ be the matrix of structure constants encoding
the multiplication by $p_1(x)=x$. 
We thus have a representation of the chiral algebra by a set of
polynomials in $x$ and 
according to an argument given for example in \DFZf, this implies that the 
constraint $W'(x)=0$ is the characteristic equation satisfied by
$C_1$: 
\eqn\IIab{ W'(x)= \det (x\un-C_1)}
In particular, if the perturbed potential $W$
is a good ``resolution" of the singular $W_0={x^{n+1}\over n+1}$, 
namely if the zeros of $W'$ 
are distinct,  then $C_1$ has distinct eigenvalues and is diagonalizable.
(An example of a non-diagonalizable case is provided by 
$W=x^6/6-x^3$; $C_1$ has a 
double eigenvalue at 0 and is not diagonalizable. The singularity has
not been resolved. ) 

Suppose now that $C_1$ is ``normalizable''. By definition, this means that
a {\it diagonal}\foot{Our insistence on this diagonal change of basis is 
due to the fact that
it must preserve the gradation by the $U(1)$ charge; when several fields
have the same gradation, a wider set of redefinitions would be conceivable; we
have not studied this systematically. } 
change of  basis in the polynomial representation makes $C_1$ normal
\eqn\IIac{ C_{1i}^{\ \ j}={\Gr_1\Gr_i\over \Gr_j}M_{1i}^{\ \ j} }
with $M_1$ normal, hence diagonalizable in an orthonormal basis 
$\psi_a^{(i)}$; in fact as all $C_i$ commute, they are all made
normal by the same change 
\eqna\IIad
$$\eqalignno{ 
C_{ij}^{\ \ k}&={\Gr_i\Gr_j\over \Gr_k}M_{ij}^{\ \ k}& \IIad a \cr
\(M_i\)_j^{\ k} &=\sum_a \Gl_i ^{(a)} \psi^{(j)}_a\psi^{(k)*}_a & \IIad b \cr}
$$
and because of the symmetry $i \leftrightarrow j$, the 
condition $p_0=1$ hence $M_0=\un$  and the orthonormality
of the $\psi$'s, one finds that the eigenvalues have the form
$\Gl_i ^{(a)}={\psi_a^{(i)}\over \psi_a^{(0)}}$ hence
\eqn\IIae{ \(M_i\)_j^{\ \ k} =\sum_a {\psi^{(i)}_a \psi^{(j)}_a\psi^{(k)*}_a
\over\psi^{(0)}_a}\ . }
Thus the eigenvalues of $C_1$, i.e. the zeros of $W'$, or the extrema
of $W$, are
\eqn\IIaf{ x_a= \rho_1 {\psi_a^{(1)}\over\psi_a^{(0)}}\ . }
%


The case where the potential involves more than one variable is easy 
to deal with, but at one point less explicit. For definiteness we
consider the case of 2 variables but the extension to any larger
number is straightforward. The assumption 
that the theory is described by a potential $W(x,y)$ amounts to saying 
that the chiral ring is represented by polynomials in the variables
$x$ and $y$, i.e. that all the matrices $C_i$ are polynomials 
in the matrices $C_0=\un$, $C_x$ and $C_y$ 
\eqnn\IIag
$$\eqalignno{
x p_i(x,y)&= C_{xi}^{\ \ j}p_j(x,y) \qquad \mod \partial_x W,\partial_y W\cr
y p_i(x,y)&= C_{yi}^{\ \ j}p_j(x,y) \qquad \mod \partial_x W,\partial_y W \ .
&\IIag\cr}$$
Thus for any extremum of the potential $x_a,y_a$, 
$\partial_x W(x_a,y_a)=\partial_y W(x_a,y_a)=0$
\eqnn\IIah
$$\eqalignno{
x_a p_i(x_a,y_a)&= C_{xi}^{\ \ j}p_j(x_a,y_a) \cr
y_a p_i(x_a,y_a)&= C_{yi}^{\ \ j}p_j(x_a,y_a) & \IIah\cr
}$$
and hence $x_a$ is an eigenvalue of $C_x$ and $y_a$ one of $C_y$ for the same 
eigenvector. If moreover $C_x$ and $C_y$ are normalizable, then 
all the $C$'s may be diagonalized in the orthonormal basis $\psi^{(i)}_a$
according to eq. \IIad{a} and as before, 
\eqnn\IIai
$$\eqalignno{
x_a& = \rho_x {\psi_a^{(x)}\over\psi_a^{(0)}}\cr
y_a& = \rho_y {\psi_a^{(y)}\over\psi_a^{(0)}}& \IIai\cr 
}$$
In contrast with the one-variable case, the reconstruction of $W$ from
$C_x$, $C_y$ is not obvious.

The appearance of algebras of the form \IIae,  generalizing 
the Verlinde formula for fusion algebras is something that has already
been encountered in association with graphs. We shall devote the next 
section to recall some facts on what we call graph algebras. 

%
%

\newsec{Graph algebras}
\subsec{The two dual algebras attached to a graph}
\noindent We present here some concepts on rings (or 
algebras) attached to graphs \refs{\Vin,\DFZ}. 
Let us consider a  graph defined by its adjacency matrix $G$, 
whose non negative entries $G_{ab}$ count the number of edges connecting
the vertex $a$ to the vertex $b$. The 
graph may possibly be oriented, and thus the matrix $G$ be non symmetric, 
but we request it to be normal. 
(Here, $G$ is real and thus $G^{\dagger}=G^t$). Clearly any
symmetric matrix (hence any adjacency matrix of an unoriented graph) is
normal. Let us denote $\psi^{(\ell)}_a$ 
the components of the orthonormal eigenvectors, where
the index $\ell$ 
labels
the eigenvector. Note that in general, $\ell$ and $a$ take an equal number 
of values (equal to the size $n$ of the $G$ matrix) but belong to different
sets. For convenience, we shall label the vertices $a$ by integers 
running from 0 to $n-1$, whereas 
$\ell$ will also
for simplicity be taken as an integer taking in particular the value 0
\foot{We depart from our previous
conventions of \DFZ, where both $a$ and $\ell$ were taking values starting 
from 1 rather than 0; this change is motivated by the mismatch with the 
degrees of polynomials or other gradings ($U(1)$ charges...) that are
natural in the problem at hand.}.
By convention, $\ell=0$ will denote the Perron-Frobenius eigenvector. We 
also {\it assume} that $G_{ab}$ possesses a ``unit" vertex $a_0$
such that 
\eqna\IIIaa
$$\eqalignno{
&(i) \quad \forall \ell,\quad
\psi^{(\ell)}_{a_0}\ne 0.  & \IIIaa a\cr
&(ii) \ \exists !\ {\rm vertex} \ f \quad {\rm such \ that } \quad  
G_{a_0,a}=\delta_{a,f} &\IIIaa b \cr
}$$
At the possible price of a relabelling of the vertices of the
graph we shall take $a_0=0$. The role of this unit vertex $0$ is clear:
the graph associated to $G$ encodes some sort of ``fusion" by the vertex $f$,
in the sense that we can write $f \times 0 = f$, $f \times a= G_{ab} \ b$.
In the particular case of $ADE$ Dynkin diagrams, $\ell$ is a 
Coxeter exponent minus 1 taking $n$ values between 0 and $h-2$, 
with $h$ the Coxeter number. 
More generally, in a variety of cases related to a Lie algebra, 
it is more natural to regard it as taking its values in a bounded 
domain of the weight lattice of this Lie algebra \DFZ.
The archetypical case is provided by the $A_{n}$ Dynkin diagram, where the 
$\psi^{(\ell)}_a$, $\ell, a=0,\cdots,n-1$ are also the matrix elements of the 
modular $S$-matrix for the $SU(2)_{n-1}$ current algebra. This is readily
seen on the celebrated Verlinde formula \EV\ written as 
\eqn\IIIa
{ M_{\ell m}^{\ \ p}=\sum_{a=0}^{n-1}{\psi^{(\ell)}_a \psi^{(m)}_a 
\psi^{(p) \,*}_a  \over \psi^{(0)}_a }\ .}
expressing the fusion coefficients in terms of $\psi$'s.  
The $A$ Dynkin diagram has the uncommon property of being self-dual in the
sense that the two sets of $a$ indices and $\ell$ labels may be identified: 
this is due to the symmetry of the $S=\psi$ matrix. 
Now, the Verlinde formula suggests to form similar sums for the other
$D$ or $E$ Dynkin diagrams, or more generally for a generic graph with a 
normal adjacency matrix. Then the two sets $\{a\}$ and
$\{\ell\}$ are no longer equivalent and there are two possible summations. 
The one carried out in \IIIa, and the dual one
\eqn\IIIb
{ N_{ab}^{\ \ c}=\sum_{\ell=0}^{n-1}{\psi^{(\ell)}_a \psi^{(\ell)}_b 
\psi^{(\ell) \,*}_c \over \psi^{(\ell)}_0 }\ .}
Contrary to the case of the $A$ Dynkin diagram, the $M$'s are not 
in general integers. In contrast, for the $D$ and $E$ Dynkin diagrams as well
as a larger class of graphs studied in \DFZ, the $N$'s are ! Moreover, for a 
subset of these graphs, -- the $A$, $D_{{\rm even}}$ and $E_{6,8}$ cases
among the Dynkin diagrams--, both $M$'s and $N$'s turn out to be non
negative. We stress that these are empirical observations and that we 
know no sufficient condition on the graph that ensures the integrality of
the $N$'s. Note that the matrices $M_{\ell}$ and $N_a$ defined
respectively by 
\eqnn\IIIc
$$\eqalignno{\(M_{\ell}\)_{m}^{\ \,p}&=M_{\ell m}^{\ \ p} \cr
\(N_a\)_b^{\ c}&=N_{ab}^{\ \ c} &\IIIc\cr}$$
satisfy an associative and commutative algebra 
\eqnn\IIId
$$\eqalignno{M_{\ell}M_{m}&= M_{\ell m}^{\ \ p}M_{p}\cr
N_a N_b &= N_{ab}^{\ \ c} N_{c} &\IIId\cr }$$
and the orthonormality condition ensures that $M_0=\un$ and $N_0=\un$ are the 
units of these algebras. 

In writing \IIIa, \IIIb, we have implicitly assumed that these summations
make sense, i.e. that no vanishing denominator occurs. Although the 
Perron-Frobenius theorem tells us that the components of the eigenvector 
with the largest eigenvalue are non-negative, it does not forbid the 
vanishing of some of these. We do know cases where either one of the
$\psi^{(\ell)}_0$ or one of the $\psi^{(0)}_a$ vanishes.
To avoid the possibility of a vanishing $\psi_0^{(\ell)}$,
we included the condition (i) \IIIaa a\ in the definition of the
unit vertex $0$ of the graph.
Moreover, a sufficient condition to avoid the vanishing of  
$\psi_a^{(0)}$'s is to suppose that the graph is connected, i.e.
there exists a path $a_1=a,a_2,...,a_p=b$ between any couple of vertices
$(a,b)$ of the graph, 
$G_{a_1 a_2} G_{a_2 a_3} ... G_{a_{p-1} a_p}\neq 0$.

We should also note that there are cases where the matrix $G$ has
degenerate  eigenvalues and there is a problem of choosing the 
appropriate combination of the corresponding eigenvectors. Such is the
case of the $D_{{2\ell}}$ Dynkin diagram: the middle exponent $2\ell-1$ is
twice degenerate and the coefficients $N_{ab}^{\ c}$ 
are integers only for a specific choice of the eigenvectors 
$\psi^{(2\ell-2)\pm}$; for $\ell$ even, this choice involves complex 
combinations of
the real eigenvectors, whence the relevance of the complex conjugation in 
eqs. \IIIa-\IIIb.

Except in the simplest case of $A$ type, no physical interpretation
of these algebras, and in particular of the integral $N$'s as 
some multiplicities is known
(see \VP, however).

%
\subsec{Subalgebras and modular invariance}
\noindent In this section, we review the connections between  
(some of) these graph algebras and fusion algebras of rational
conformal field theories. This is not in the main stream of 
our paper, but we include it here to illustrate some
crossrelations between these topics and to show that some of
the considerations of \LW\ may be extended. Whenever 
all the $M$'s and $N$'s are non negative (case referred as ``type I" in the
\DFZ), one may find a subalgebra of the graph algebra
\IIIb, i.e. a stable subset $T$ of vertices of the graph, 
which encodes the {\it fusion rules} of the 
underlying WZW
theory. Namely, in the cases considered in \DFZ, 
each of the graphs forms the target space for an integrable
lattice model whose continuum limit is described by a coset c.f.t.
$G_{k-1}\times G_1/G_k$ and is in  correspondence with a modular invariant
of the relevant $G_k$ WZW theory. 
The $\ell$ labels indexing the algebra $M$ are in 
correspondence with integrable weights of the $G_k$ WZW
model at some fixed level $k$.  On the other hand the modular invariants 
of the WZW theories at a given level can be of two forms: the 
``block-diagonal" ones
formed by a sum of absolute squares of sums of WZW characters $\chi_{\ell}$ 
(the so--called
``extended" characters), and the ``twisted" ones, obtained by combining
left and right blocks of the preceding class in a non--diagonal way.
Concentrating on the block-diagonal invariants (also called type I
in \DFZ) \foot{This distinction between type I and non-type I graphs and/or
modular invariants seems to have multiple aspects, as testified by the
existence or non-existence of a flat connection on the space of paths on 
the graph \OEK; see also the end of Appendix A.}, we see that they are 
characterized by a partition $I_1,I_2,...,I_p$
of the set of labels of representations
which form the theory, the modular invariant reading
$$ Z= \sum_{i=1}^p | \sum_{\ell \in I_i} \chi_{\ell} |^2 .$$
We found that the graph subalgebra coincided in all known cases with
the fusion rules for
the primary states of the ``extended" symmetry, 
whose characters are the blocks 
\eqn\III
{\chi_{I_k} = \sum_{\ell \in I_k} \chi_{\ell}\ . }
More precisely, from the data of the graph ring and and of its 
subring associated with the stable subset $T$ of
vertices, we can define the equivalence relation $\simeq$ 
on the eigenvalue labels \BI 
\eqn\eclas{
 \ell \simeq m \qquad {\rm iff} \ \ \ \sum_{a \in T} \psi_a^{(\ell)}
\psi_a^{(m) \, *} \neq 0\ .}
The equivalence classes form the desired partition of the set of eigenvalue labels
into blocks $I_1,...,I_p$. The latter are in one to one correspondence with 
the elements of $T$, and we relabel them $I_a, \ a \in T$.
The corresponding $S$ matrix of modular transformation
%
%
is given by 
\eqn\IIIag{
S_a^{\ b} = {\psi_a^{(\ell)} \over (\sum_{c \in T} |\psi_c^{(\ell)}|^2)^{1/2}}
,\ \ {\rm independent \ on }\  \ell \in I_b\ .}
Like any fusion algebra of a rational conformal field theory, 
the subalgebra with fusion coefficients
$N_{a b}^{\ \ c}, a,b,c \in T$ is self--dual, due to the symmetry of $S$.
The dual $M_{ab}^{\ \ c}$ can also be expressed in terms of the original 
dual algebra $M_{\ell m}^{\ \ n}$, but in a less straightforward way.
In particular its one dimensional representations have a simple
realization in terms of the one dimensional representations of 
$M_{\ell m}^{\ \ n}$. The latter are of the form
\eqn\IIIah{\tilde p^a_{\ell}={\psi_a^{(\ell)} \over \psi_a^{(0)}} }
and satisfy $\tilde p^a_{\ell}\tilde p^a_{m}=M_{\ell m}^{\ \ n}
\tilde p^a_{n}$,
for any vertex $a$. The corresponding one dimensional representations
of the self--dual subalgebra read
\eqn\IIIai{\Pi_a^{(b)}={S_a^{\ b} \over S_a^{\ 0}} \ \ \ a,b \in T\ ,}
and one finds, using the various relations between the $S$'s and the $\psi$'s
\eqn\IIIaj{\Pi_a^{(b)}=\sum_{\ell \in I_b} \bigg[ \sum_{c \in T}
|\psi_c^{(0)}|^2
\sum_{c \in T}|\psi_c^{(\ell)}|^2 \bigg]^{1 \over 2}\ \tilde p^a_{\ell}\ ,}
for any $a,b \ \in T$.

We should stress that all these considerations are empirical and based on 
a case by case examination of all the type I cases pertaining to 
$G= SU(2)$ or $SU(3)$ (the classification of modular invariants for the 
latter case has been completed in \GAN).



Let us illustrate this on an example. We consider the $E_6$ diagram
of Table I, and the subalgebra of the graph algebra formed by the 
end-point vertices, $T=\{0,4,5\}$.
It is isomorphic to the fusion algebra of the Ising model, 
(known to be that of the blocks of the $E_6$ theory),  
upon identification of $0\equiv Id$, $4\equiv  \epsilon$
and $5\equiv \sigma$, respectively the identity, spin and
energy conformal blocks of the Ising model.
The eigenvalues of the adjacency matrix $[E_6]_{ab}$ are
labelled by the Coxeter exponents shifted by $-1$ 
$$\beta^{(\ell)}= 2 \cos \pi {\ell+1 \over 12} \quad \ell=0,3,4,6,7,10.$$ 
Applying eq.\eclas, we find the corresponding equivalence classes
of the set of exponents
$$I_0=\{0,6\} \quad I_4=\{4,10\} \quad I_5=\{3,7\},$$
and the associated modular invariant
\eqn\esix{
Z_{E_6}=|\chi_0+\chi_6|^2+|\chi_4+\chi_{10}|^2+|\chi_3+\chi_7|^2\ .}
Moreover we get the one dimensional representations of the self dual
subalgebra
\eqnn\IIIak
$$\eqalignno{\Pi_a^{(0)}&= {1 \over {3+\sqrt{3}}}\tilde p^a_{0} +
 {1 \over \sqrt{6}} \tilde p^a_{6} \cr
\Pi_a^{(4)}&={1 \over \sqrt{6}} \tilde p^a_{4} + {1 \over {3+\sqrt{3}}}
\tilde p^a_{10} 
& \IIIak \cr
\Pi_a^{(5)}&={1 \over \sqrt{2(3+\sqrt{3})}}\[ \tilde p^a_{3}+
\tilde p^a_{7}\]\ . \cr}
$$
%



\subsec{\Che resolution of the $ADE$ singularities}
\noindent Let us show how the considerations of Sec. 2  apply 
to the case considered by Lerche and Warner, namely the perturbations
of the $ADE$ potentials by their least relevant operator. In the $A_n$ case, 
the corresponding potential is nothing else than the \Che polynomial
of the first kind ($T_n(2\cos\Gth)=2 \cos n\Gth$) : $W(x)={1\over n+1} 
T_{n+1}(x)$; 
the basis of the chiral ring derived from the flat 
coordinates is provided by the \Che polynomials of the second kind 
($U_l(2\cos \Gth)= {\sin (l+1)\Gth\over \sin \Gth}$), 
$p_{\ell}(x)=U_{\ell-1}(x)$, $\ell=1, \cdots, n$, and the chiral algebra
that they satisfy is just the $SU(2)_{k=n-1}$ fusion algebra. The 
potential $W(x)={1\over n+1} T_{n+1}(x)$ 
is the fusion potential that encodes these fusion rules \refs{\DGe,\DFZf}. 

The other cases $D$ and $E$ have also been discussed \LW. By inspection, one
finds that 
\item{i)} the corresponding matrices $C_x$ and $C_y$ can be made normal
by a diagonal redefinition of the basis;
\item{ii)} the dual algebra has among its generators the incidence matrix
of the Dynkin diagram, or equivalently the $\psi^{(\ell)}_a$ are the
eigenvectors of the Cartan matrix of the $D$ or $E$ Lie algebra and according
to the discussion of Sec. 2, $p_{l}(x_a,y_a) \propto \psi^{(l)}_a/
\psi^{(0)}_a$;
\item{iii)} the pattern of extrema of the potential in the $x-y$ plane 
reproduces the shape of the Dynkin diagram.

Then the previous discussion applies: in the ``good'' cases $D_{{\rm even}}$,
$E_6$ and $E_8$, one can find linear combinations of the polynomials 
$p_l(x,y)$ that generate a {\bf subring} of the chiral ring isomorphic
to the fusion ring of the corresponding $SU(2)$ modular invariant. Let us
illustrate this again on the case of $E_6$. We start from the deformed
$E_6$ potential $W$ given in \DVV\ and recalled for convenience in Appendix B.
The polynomials $p_i(x,y,\tt)= -{\partial\over \partial t_i}W$ 
form a ring (modulo $\partial_x W$,$\partial_y W$) with structure 
constants derived from the free energy given in \KTS. The potential 
$W$ and the $p$'s are quasihomogeneous polynomials of $x$ (of degree 4), 
$y$ (degree 3) and $t_i$ (degree $12-i$).

If only $t_{10}=t$, the coupling to the least relevant operator, is  
non vanishing, the polynomials $p_i$ reduce to 
$p_0=1$, $p_3=y$, $p_4=x-\oh t^2$, 
$p_6=y^2-tx+{1\over 6}t^3$, $p_7= xy-t^2y$ and $p_{10}=xy^2-
{3\over 2}t^2 y^2+{1\over 3}t^3x$. 
After a change of scale
\catcode`\@=11
\def\Eqalign#1{\null\,\vcenter{\openup\jot\m@th\ialign{
\strut\hfil$\displaystyle{##}$&$\displaystyle{{}##}$\hfil
&&\qquad\strut\hfil$\displaystyle{##}$&$\displaystyle{{}##}$
\hfil\crcr#1\crcr}}\,}   \catcode`\@=12
$$\Eqalign{
\tilde p_0&= p_0  &\tilde p_3&={\sqrt{3}\over t^{{3\over 2}}} p_3 \cr
\tilde p_4&={\sqrt{6}\over t^2} p_4 &\tilde p_6&={3\sqrt{2}\over t^3} p_6 \cr
\tilde p_7&={6\over t^{{7\over 2}}} p_7 &\tilde p_{10}&={6\sqrt{3}\over t^5} p_{10}\cr }$$
the dimensionless $\tilde p$'s have structure constants given by the algebra
dual to the one generated by the $E_6$ Dynkin diagram, in the sense
of Sec. 3.1. In other words, 
$$\tilde p_i \tilde p_j= M_{ij}^{\ \ k}\tilde p_k\ .$$
To make contact with the fusion algebra of the underlying $SU(2)$ model
we finally form linear combinations of the $\tilde p$'s according to 
eq. \IIIak, and we find that the polynomials $\Pi_0$, $\Pi_4$ and 
$\Pi_5$ form a subring isomorphic to the fusion ring of the theory \esix.

The property of normalizability enjoyed by the chiral ring of the 
\Che resolution of the $ADE$ singularities has prompted us to 
systematically examine what are the normalizable deformations of these
cases. This will be our endeavour in the next section.

%
\newsec{Normalizable deformations of the $ADE$ singularities}
\noindent This section is a catalog of the normalizable 
deformations of the $ADE$ potentials by a single non vanishing 
parameter $t_i$ (see section 6 for a discussion of this restriction). 
Because the potential is a  quasihomogeneous polynomial
of the variable(s) $x$ (and possibly  $y$) and of this  parameter, $t_i$ 
may be rescaled to the value 1. This will be assumed in the following.

\subsec{$A_n$}
\noindent We first examine the $A_n$ deformed potential $W(x,t_0,\cdots,
 t_{n-1})$ of \DVV. The matrix $C_1$ in that case reads
\eqn\IVa{C_1= \pmatrix{ 0 & 1 & 0 & \cdots& \cr
	 	       t_{n-1}& 0 & 1 &0 & &\cr
		       t_{n-2}& t_{n-1}& 0 & 1& \ddots \cr
		       \vdots  & \ddots & \ddots & \ddots & \cr
		       t_1 & \cdots& t_{n-2}& t_{n-1}& 0\cr } \ }
and the potential $W$ is reconstructed from \IIab\ by one quadrature.
We assume that only $t_{p+1}=t$ is non vanishing.
Then $C_1$ reads
\eqn\IVa{C_1= \pmatrix{ 0 & 1 & 0 & \cdots&  &\cdots &0\cr
	 	         & 0 & 1 &0 & &  & 0\cr
		         &  & 0 & 1& \ddots &  & \cr
		        0 & &  & \ddots &\ddots&\ddots&  \cr
		        t & 0 & \cdots&  &  &\ddots  & 0\cr 
                         & \ddots & \ddots&  & &  0&1 \cr                      
                        0 & \cdots &t &0& & \cdots &0\cr} \ }
where the diagonal of $t$'s starts in position $(n-p,1)$ on the 
matrix.
We look for a diagonal change of basis $P=\diag(\rho_1,...,\rho_n)$,
such that ${\tilde {C_1}}=P^{-1} C_1 P$ is normal.
The normality condition imposes constraints on the $\rho$'s, 
which are all expressed in terms of $\sigma_i=\rho_i^{-2}$.
For all values of $p$, we get
\eqn\allts{ {\sigma_{n-p} \over \sigma_1}={\sigma_{n-p+1}\over
\sigma_{2}}=\cdots={\sigma_n \over \sigma_{p+1}}.}
According to the respective positions of $p$ and $n \over 2$,
we find for $p \geq 1$
$$\eqalign{
(i) \ \ \ p<{n \over 2} & \cr
\qquad 1 \leq j \leq p+1 &\qquad {\sigma_j \over \sigma_{j+1}}=
j{\sigma_1 \over \sigma_2} \cr
\qquad p+2 \leq j \leq n-p-1 &\qquad {\sigma_j \over \sigma_{j+1}}=
(p+1){\sigma_1 \over \sigma_2} \cr
\qquad n-p \leq j \leq n-1 &\qquad {\sigma_j \over \sigma_{j+1}}=
(n-j){\sigma_1 \over \sigma_2} \cr
(ii) \ \ \ p \geq {n \over 2} & \cr
\qquad 1 \leq j \leq n-p-1 &\qquad {\sigma_j \over \sigma_{j+1}}=
j{\sigma_1 \over \sigma_2} \cr
\qquad n-p \leq j \leq p+1 &\qquad {\sigma_j \over \sigma_{j+1}}=
(n-p-1){\sigma_1 \over \sigma_2} \cr
\qquad p+2 \leq j \leq n-1 &\qquad {\sigma_j \over \sigma_{j+1}}=
(n-j){\sigma_1 \over \sigma_2} \cr}$$
{}From $(i)$ and \allts\ we get 
$${\sigma_{n-p} \over \sigma_{n-p+1}}={\sigma_1 \over \sigma_2}
=p{\sigma_1 \over \sigma_2},$$
possible only if $p=1$.
Analogously, from $(ii)$ and \allts\ we get
$${\sigma_{n-p} \over \sigma_{n-p+1}}={\sigma_1 \over \sigma_2}
=(n-p-1){\sigma_1 \over \sigma_2},$$
possible only if $p=n-2$.
This leaves us with the three cases 
$$\eqalign{
1)\,\ \ \ \ \  \ p=0 \ :&\ \sigma_k=2^{-2(k-1)/n} \qquad k=1,2,...,n. \cr
2)\,\ \ \ \ \  \ p=1 \ :&\ \sigma_k=2.(2t)^{-2(k-1)/(n-1)}\ \  k=2,3,...,n-1, \cr
&\ \sigma_1=1 \qquad \sigma_n=t^{-2} \ . \cr
3) \ p=n-2 \ :&\ \sigma_k=t^{-(k-1)} \qquad k=1,2,...,n. \cr}$$

We conclude that the only normalizable cases with a single non-vanishing 
$t$ parameter are the three cases $t_1\ne 0$, $t_2\ne 0$ and $t_{n-1}\ne 0$. 
What is most striking is that these three cases have been identified 
as integrable deformations \MW: the perturbation by the most relevant
operator has long been recognized as integrable 
(\refs{\LVW,\FLMW\LWnp\CVtat{--}\FIii}), 
the one by $t_2$ is discussed in \refs{\FMVW,\FLMW} 
while the case of $t_{n-1}$ is treated in \LW\ and \FIi.	 

There are several things that can be done on these normalizable cases:
\item{a)}
{}From the diagonalization of the matrix $\tilde C_1=\rho_1 M_1$ in an 
orthonormal basis,
we can construct the dual algebra $N_{ab}^{\ \ c}$. In all these cases, 
it exists (because 
all $\psi^{(\ell)}_0 \ne 0$) and it leads to {\it non-negative} integers ! 
Each of these can be regarded as the adjacency matrix of a (possibly 
disconnected) graph; 
\item{b)} 
We can also determine the extrema of the potential $W(x,\tt)$, i.e.
both the location of the extrema $x_a$ (on the real line or in the complex 
plane) and the value of $W$ at this $x_a$. We find that the location
of the $x_a$ follows the pattern of vertices of one of the graphs 
of the dual algebra, call it $N_{fb}^{\ \ c}$; consequently, it seems 
natural to link the extrema $x_a$ by edges of the graph of $N_f$. 
As for the extremal values of $W$, they are such that for two
extrema $x_a$ and $x_b$ linked as just explained, 
$|W(x_a)-W(x_b)|$ takes only one value $|\Delta W|$. The interpretation
\FMVW\ is that the link exists between the ground states $a$ and $b$ of the
potential if and only if there is a ``fundamental"
 soliton interpolating between 
them, and the mass of this soliton is just given by $|\Delta W|$.

These features are apparent on the graphs tabulated in Table I. 
 We comment briefly the results.\par
\item{$\bullet$} 
For the perturbation by $t_{n-1}$, the \Che resolution discussed 
before, the extrema lie at $x_a= 2\cos {\pi (a+1) \over n+1}$, $a=0,
\cdots, n-1$, and the value of $W$ at these points is 
$W_a= {2(-1)^{a+1}\over n+1}$. The graph encoding $C_x$ as well as
one of its dual is the $A_n$ Dynkin diagram.\par
\item{$\bullet$} 
For the perturbation by $x^2$, $W= {x^{n+1}\over n+1}- \oh x^2$, one finds 
extrema at $x_0=0$ and $x_a= \exp \(2i\pi {a-1\over n-1}\)$, 
$a=1,\cdots, n-1$; they form a centered $(n-1)$--gon in the complex plane, 
like the corresponding values of $W$: $0$, ${n-1\over 2(n+1)}
\exp \(4i\pi {a-1\over n-1}\)$. The graph of the dual $N_{1a}^{\ \ b}$ 
has the daisy shape depicted in the third column of Table I. \par
\item{$\bullet$} 
For the perturbation by $x$, the results are similar, with a non-centered 
{\it oriented} polygon: $x_a=\exp \(2i\pi {a\over n-1}\)$, 
$W_a={-n\over n+1} \exp \(2i\pi {a\over n-1}\)$, $a=0,\cdots, n-1$. 
On table I, only the graph associated with $N_1$ has been drawn, but the
other $N$'s would connect other pairs of ground states, corresponding to
the other, non fundamental, solitons.


\subsec{$D_{n+2}$ }
\noindent It may be useful to first recall that the $D_{n+2}$ perturbed
potential may be obtained from the $A_{2n+1}$ one by an orbifold 
procedure. We devote Appendix A to 
a review of this construction and of various properties of the 
$D$ potential and free energy, including a curious positivity property
of the coefficients of $F$ for $D_{{\rm even}}$. 
The bottom line is that the $D_{n+2}$ potential involves a new variable $y$, 
and after inserting an extra deformation parameter $\tau$ coupled to $y$, 
it reads
\eqnn\IVaa
$$\eqalignno{W_{D_{n+2}}(x,y;t_0,t_2,\cdots,t_{2n},\tau)
&= W_{A_{2n+1}}(x'=\sqrt{x}, 
t_0,0,t_2,\cdots,t_{2n}) + \oh x y^2 +\tau y\ . &\ \cr
&= {x^{n+1}\over 2(n+1)} + \cdots  &\IVaa \cr}$$
The free energy $F$ of the $D$ models may also be determined fairly
explicitly in terms of the $A$ one. One finds that 
\eqn\IVab{ F_{D_{n+2}}(t_0,t_2,\cdots, t_{2n},\tau)=F_{A_{2n+1}}(t_0,0,t_2,\cdots, t_{2n})
-{\tau^2\over 2} \Phi(t_0,\cdots,t_{2n})}
where the expression of $\Phi$ is given in Appendix A. 

{}From the orbifold connection between the $D_{n+2}$ and the $A_{2n+1}$ 
cases, it seems reasonable to expect the
perturbations by the least relevant and the most relevant operator
to be i) integrable, ii) normalizable; the former observation has been 
made in \FLMW\ and \NW, for the most relevant (and $n$ even), and
\LW, for the least relevant; as for normalizability, it is readily checked
for these two 	perturbations and one also finds that the 
perturbation by $\tau$ is normalizable. The proof goes as follows.  

The matrix encoding the multiplication by $p_1(x)$ is given by
\eqnn\cdIII
$$\eqalignno{ (C_1)_i^{\ j} &= {\partial\over \partial t_2}
{\partial\over \partial t_i}{\partial\over \partial t_k} \(F_{A_{2n+1}}
 - {1 \over 2 }\tau^2 \Phi (t.)\)
\eta^{j,k} \cr 
C_1 &= \pmatrix{ & & & 0 \cr
			  & A & & \vdots \cr
			  & & & - \tau \cr
			  \tau & 0 & \cdots & -t_{2n} } & \cdIII\cr } $$
where $A$ is a $(n+1)\times (n+1)$ matrix.
which can be expressed in terms
of the matrix $\widehat{C}_1$ encoding multiplication by $x$ in the
$A_{2n+1}$ model.  The relation is as follow,
\eqn\cdVI{ (A)_i^{\ j}=(\widehat{C}_1^2)_{2i}^{\ 2j}-t_{2n}\delta_{ij}  , \  
i=0,...,n.}
Looking at specific perturbations we now get:
\item{$\star$} : $\tau$-perturbation:
The matrix $C_1$ simplifies dramatically and is,
\eqn\cdVII{C_1= \pmatrix{0 & 1 & 0 & \cdots & & 0 \cr
			 0 & 0 & 1 & 0 & \cdots & 0 \cr
			 \vdots & \ddots & & & \ddots & \vdots \cr
			 0 & \cdots & & 0 & 1 & 0 \cr
			 0 & \cdots & \cdots & 0 & 0 & -\tau \cr
			 \tau & \cdots & & & 0 & 0 } \ .}
We see that this is a cycle and it is obviously normalizable.

\item{$\star$} : $t_i$-perturbation.
The matrix $C_1$ has the following form,
\eqn\cdVIII{C_1= \pmatrix{ & & & 0 \cr
			   & A & & \vdots \cr
			   & & & \cr
			   0 & \cdots & & -t_{2n} } \ . }
Obviously if in that case $A$ is normalizable, the same property
holds for $C_1$, therefore the normalizable cases 
in $A_{2n+1}$ with $t_{{\rm odd}}=0$ will also be in this case. 
It is likely that these are the only normalizable perturbations with a
single non-vanishing $t_i$, although we have no complete proof. There is, 
however, the possibility to mix the two operators of same degree, and 
for $D_6$, for example, one finds that the operators coupled to 
$t_n\pm i \tau$ are normalizable (see below sec. 5.2). 

\noindent We thus discuss in turn the perturbations
\item{$\bullet$}  by $t_{2n}$. This is the \Che perturbation discussed 
in \LW\ and \FIi. 
The minima in the $x$-$y$ plane build up the shape of the Dynkin
diagram, while $W$ takes only two values. 
It is still true that the locations of the extrema are related to
the eigenvectors of the adjacency matrix of the Dynkin diagram $D_{n+2}$
as in \IIai. If we want to reconstruct the whole $M_{ab}^{\ \ c}$ dual
algebra, however, we have to distinguish the cases of even and odd $n$. 
For even $n$, one chooses for the vertex $0$ the end of the longest leg of
the diagram, since all the components $\psi^{(\ell)}_0$ are non vanishing
(after some judicious choice of linear combinations of the eigenvectors
pertaining to the same eigenvalue).
For $n$ odd, in contrast, one has to take rather the end of a short
leg to have a well-defined expression and then the $N_{ab}^{\ \ c}$'s are not 
all non negative, as recalled in sec. 3.1.
\item{$\bullet$}  by $\tau $. 
The potential $W= {x^{n+1}\over 2(n+1)}+\oh xy^2-y$
has minima at $x_a \propto \exp {2i\pi a\over n+2}$, $a=0,\cdots,n-1$ and
takes there the values $W_a \propto x_a^{-1}$ (the overall factors or 
phases have been discarded in Table I). The multiplication by $x$ 
yields a matrix of cyclic permutation, and the dual $N_1$ has the
same form: in Table I, the links of the graph should be oriented. 
The integrability of that case is, 
to the best of our knowledge, not established. 
\item{$\bullet$}  by $t_{2}$. The potential $W= {x^{n+1}\over 2(n+1)}+\oh xy^2-x$ has 
minima at $x_a =2^{1/n} \exp{2i\pi a \over n}$, $y_a=0$ for
$a=0,1,...,n-1$ and $x=0$, $y=\pm \sqrt{2}$, and the values 
of the potential are $W(x_a,y_a)=-{n 2^{1/n} \over n+1}
\exp{2i\pi a \over n}$, $W(0,\pm \sqrt{2})=0$.
For even $n$, the ring generators are normal in the basis 
$$1,x,x^2,...,x^{n/2-1},
{x^{n\over 2} \pm \omega y \over \sqrt{2}},x^{n/2+1},..,x^{n-1},
x^n-1, $$
for any $\omega$ such that $\omega^4=1$. The dual algebra generator
is a cyclic permutation of $n$ vertices together with the exchange
of the two remaining ones. For $n$ odd, the ring generators are normal
in the basis $1,x,\cdots, x^{n-1}, x^n-1,y$. The dual ring
cannot be constructed, due to a failure of condition \IIIaa{a}.
For even $n$, the perturbation has been argued to be integrable, 
as the most relevant one \NW. 

\subsec{$E_6$}
\noindent The only normalizable perturbations with a single non-vanishing
$t_i$ are:
\item{$\bullet$} by $t_{10}$. This is the \Che perturbation, with the extrema
of the potential at either $y=0$, $x=\pm {1\over 2\sqrt{3}}$, $W= \mp {1\over
36\sqrt{3}}$, 
or $x=1 \pm {1\over 2\sqrt{3}}$, $y=\Ge \sqrt{1\pm {1\over \sqrt{3}}}$,
($\Ge=\pm 1$),  
$W= \mp {1\over 36\sqrt{3}}$. The matrix $N_1$ is the adjacency 
matrix of the Dynkin diagram. 
\item{$\bullet$} by $t_{7}$. This is an interesting case 
where the perturbation couples the two variables $x$ and $y$ : 
$W= {x^3\over 3}+{y^4\over 4}- xy$. The extrema occur either at 
$x_0=y_0=0$, $W_0=0$ or at $x= e^{6i\pi (a-1)/5}, y= e^{2i\pi (a-1)/5}$,
$W_a=-{5\over 12} e^{-2i\pi (a-1)/5}$, $a=1\cdots, 5$. The dual
$N_5$ is the adjacency matrix of a 
daisy graph (like in the case of $A_6$ perturbed by $t_2$).
\item{$\bullet$} by $t_{4},t_{6}$. In this case and the next, we
allow two different $t$'s to be non vanishing, in apparent 
contradiction to our previous assumption. This is because the two
variables $x$ and $y$ are in fact uncoupled, and 
we are dealing with the tensor 
product of a $A_2$ $x^3$-potential perturbed by $x$ and a $A_3$
$y^4$-potential perturbed by $y^2$. As before the perturbation 
parameters $t_4,t_6$ may be absorbed into a redefinition of $x$ and $y$ and
we choose them equal to 1. The extrema lie at $x=0, 1$ and $y=0, 
\pm \sqrt{2}$, $W=\pm 2/3$ and twice $\pm 5/3$. 
\item{$\bullet$} by $t_{3},t_{4}$. The extrema are at $x=\pm 1$, 
$y=\exp 2i a\pi/3$, $a=0,1,2$ with $W$ taking six values in the plane.

\noindent 
The integrability of the case $t_{10}$ has been discussed in \LW\ and \FIi. 
It would be quite interesting to 
find a conserved quantity or any other evidence of integrability
in the case of the $t_{7}$ perturbation.

\subsec{$E_7$} 
The expression of the perturbed potential and free
energy may be found in \KTS\ (with a little misprint corrected in our 
appendix B)\foot{We are grateful to A. Klemm for a communication on this 
subject}. The only normalizable perturbations by a single
non vanishing flat coordinate are
\item{$\bullet$} by $t_{16}$. This \Che resolution, already discussed in \LW, 
leads to a dual algebra that involves signs but is well defined. The 
generator $N_1$ is the adjacency matrix of the 
$E_7$ Dynkin diagram. The extrema of $W$ take place
at points that also reproduce that diagram, (see Table I) 
and $W$ takes only two values,
(with the conventions of \KTS) $W=\pm {1\over 2\,.\, 3^{10}}$ (4 times $+$, 
3 times $-$).
\item{$\bullet$} by $t_{10}$: The potential $W={x^3\over 3}+{xy^3}-xy$
has extrema that lie in the $x$ and $y$ complex planes, making the picture
more difficult to read. Also, all the $\psi_a$ have some 
vanishing component, making the $N$ algebra ill-defined. 
Accordingly, there is no corresponding entry in Table I. 
The integrability of that case is not known. 

\subsec{$E_8$}
\noindent The parametrization of the perturbed potential by flat
coordinates may be found in \KTS. The normalizable cases are 
perturbations by
\item{$\bullet$} by $t_{28}$: 
Extrema are at $y_a=1+{1\over \sqrt{15}}{\psi^{(1)}_a\over \psi_a^{(0)}}$,
$x_a= y_a-{4\over 5}+{1\over {15}}{\psi^{(2)}_a\over \psi_a^{(0)}}$,
where $\psi^{(0,1,2)}$ are the three eigenvectors of the $E_8$
adjacency matrix pertaining to the eigenvalues 
$2\cos {\pi\over 30}(1,7,11)$. 
The corresponding critical values of $W$ take only two 
values,$ W= \pm {1\over 20250 \sqrt{5}}$. Once again, the 
extrema display nicely in the $x$-$y$ plane the shape of the
$E_8$ Dynkin diagram. The adjacency matrix of the latter is 
reproduced by the $N_1$ matrix. 
\item{$\bullet$} by $t_{16}$:  The extrema lie at the origin, with $W_0=0$
and at the seventh roots of unity, $x_a=e^{2i\pi(a-1)/7}$, $y_a=x_a^2$, 
$W_a=-{7\over 15} x_a^3$, $a=1,\cdots,7$. 
These extrema and the resulting $N_7$ graph are again like in the case 
of $A_8$ perturbed by $t_2$. 
\item{$\bullet$} by $t_{18}, t_{10}$; 
\item{$\bullet$} by $t_{12}, t_{10}$;
\item{$\bullet$} by $t_{10}, t_6$: these last three cases correspond to 
decoupled cases  $A_4 \otimes A_6$. Their extrema and graphs are thus 
obtained as tensor products of the $A$ cases discussed above.

\newsec{Non $ADE$ cases}  
\subsec{The $SU(N)$ cases}
\noindent  The study of effective Landau--Ginzburg theories 
beyond the $ADE$ potentials becomes more delicate. 
The main difficulty is the appearance of {\it modules} in the
singularities, i.e. dimensionless
parameters decorating the potentials. The simplest example of
a module is provided by the $P_8$ singularity of ref \Arn, with a
potential 
$$x^3+y^3+z^3 +a xyz,$$
where the dimensionless parameter $a$ is the module of the singularity \VeW.
However, Gepner \DGe\ found some geometrical potentials for the
fusion rings of the $SU(N)$ WZW theories at level $k$, best expressed
through their generating function
\eqn\gensuN{\eqalign{
\sum_{m \geq 0} t^m  W_m^{(N)}&(x_1,x_2,...,x_{N-1})= \cr
-\log\big(&1-tx_1+t^2 x_2-t^3 x_3+...+(-1)^{N-1} t^{N-1}x_{N-1}+(-1)^N t^N
\big). \cr}}
The fusion ring is then the polynomial ring $\IC[x_1,...,x_{N-1}]/\{
\partial_{x_i} W_m^{(N)} \}$, with $m=k+N$. 
The ring basis corresponding to the 
integrable weights of $\widehat{SU(N)_k}$, 
$(\Gl_1,...,\Gl_{N-1})$, $\Gl_i \geq 0$, $\sum \Gl_i \leq k$
is formed by generalized \Che polynomials.
It was argued that this is just a particular perturbation of the 
chiral ring of a $N=2$ Landau--Ginzburg theory of $N-1$ superfields
$\Phi_1$,...,$\Phi_{N-1}$, with a quasi--homogeneous potential $w_m^{(N)}$, 
generated by
\eqn\uvN{ \sum_{m \geq 0} t^m w_m^{(N)}(\Phi_1,...,\Phi_{N-1})=
-\log\big(1-t\Phi_1+t^2 \Phi_2-...+(-1)^{N-1}t^{N-1} \Phi_{N-1}\big).}
There is now a bulk of evidence \refs{\VWLM,\LVW,\FIii,\Wit,\DFY}
that these Landau--Ginzburg theories 
describe the $N=2$ superconformal Kazama--Suzuki models \KZ\ based on 
the cosets ${SU(N)_k\times SO(2(N-1))_1\over SU(N-1)_{k+1}\times U(1)}$.
The potential $w_{k+N}^{(N)}$ is a quasi--homogeneous function
of degree $k+N$, if we assign the degree $j$ to the field $\Phi_j$.
The ``\Che" perturbation reproducing the fusion ring of
$\widehat{SU(N)_k}$ is therefore a perturbation by the degree $k$
operator corresponding to the weight $(k,0,...,0)$ (see Table I, 
where $t_{k,0,\cdots, 0}$ is denoted by $t_k$). 
The task of computing flat coordinates for generic perturbations of 
these potentials $w_m^{(N)}$ is formidable. However it was carried out in
one special case, corresponding to $SU(3)_3$ \KTS.
The appearance of modules, i.e. of coupling parameters with negative
or zero dimension
is clear from inspection of the possible degrees of operators in 
a generic perturbed theory.
Let ${\cal U}_{\Gl_1,..,\Gl_{N-1}}(\Phi_1,...,\Phi_{N-1})$
denote the ring basis element (generalized \Che
polynomial) with weight $(\Gl_1,...,\Gl_{N-1})$. 
It behaves like 
$${\cal U}_{\Gl_1,..,\Gl_{N-1}}(\Phi_1,...,\Phi_{N-1})
=\Phi_1^{\Gl_1} \Phi_2^{\Gl_2} \cdots \Phi_{N-1}^{\Gl_{N-1}}+ \cdots$$ 
hence its degree is $\Gl_1+2 \Gl_2+ \cdots +(N-1)\Gl_{N-1}$, and can 
go up to $(N-1)k$: it can become larger than the degree of the 
attached potential, $k+N$, as soon as $k \geq 2$ for $N > 3$,
or $k\geq 3$ for $N=3$. 
Hence the perturbations by such operators
will have zero-- (marginal operators) or negative-- (irrelevant operators) 
dimension coupling constants to preserve
the quasi--homogeneity of the perturbed potential. In all cases,
these will enable to construct dimensionless couplings, whence
modules. This explains also how the cases $SU(3)_1$ and $SU(3)_2$
avoid the problem, being just part of the $A,D,E$
classification of singularities without modules (resp. $A_3$
and $D_6$, see next subsection for a detailed study), as well as 
$SU(N)_1$ ($A_N$).

Although the complete expression for perturbed potentials 
is not known, we
can consider some special perturbations which are relevant
enough to avoid the problem of modules, such as the ``\Che" perturbation
for instance. In the remainder of this section we will concentrate on the 
$SU(3)$ case at levels $k \geq 3$. 
The generating function for the potentials $w_{k+3}^{(3)}$
is easily recognized as a special form of the $SU(2)$ \Che
potential generating function, i.e.
$$\sum_{m \geq 0} t^m w_{m}^{(3)}(x,y)= \sum_{m \geq 0}
t^m y^{m/2} w_m^{(2)}(x/\sqrt{y}),$$
where $w_m^{(2)}(x)=T_m(x)/m$, $T_m$ the 	\Che polynomial
of the first kind.
This enables us to study the most relevant perturbations of the 
conformal theory, by the operators ${\cal U}_{1,0}=x+...$,
${\cal U}_{2,0}=x^2-y+...$, ${\cal U}_{0,1}=y+...$, with respective 
degrees $1$, $2$, $2$, and for which the perturbed potentials read
$$\eqalign{
x:&\qquad w= y^{n/2} {T_n(x/\sqrt{y}) \over n} -t_{1,0} x \cr
y:&\qquad w= y^{n/2} {T_n(x/\sqrt{y}) \over n} -t_{0,1} y \cr
x^2-y:&\qquad w= y^{n/2} {T_n(x/\sqrt{y}) \over n} -t_{2,0}(x^2-y).
\cr}$$
For these 
perturbations, we worked out the 
perturbed ring and found that only ${\cal U}_{1,0}=x$ and 
${\cal U}_{0,1}=y$ were normalizable, together with the \Che
perturbation by ${\cal U}_{k,0}$. On the other hand, we computed the
extrema of the potential, and found striking similarities between
the dual of the normalized ring and the positions of the extrema
and values taken by the potential at those.  
The results are collected in pictorial form on table II. 
To comment briefly, the ``\Che'' perturbation by $t_{n-3}$ leads to
a set of extrema in the $x$-plane 
inside a three-cusp hypocycloid, a deformed version
of the Weyl chamber of level $n-3$. The potential takes three 
possible values according to the triality of the ground state, on the
vertices of an equilateral triangle.  The perturbation by $t_1$ coupled
to $x$ leads to two possible pictures depending on the parity of $n$, 
because of parity properties of \Che polynomials. 
For $n$ even, the extrema in the $x$-plane 
lie at the vertices of ${n-2\over 2}$ concentric regular $(n-1)$-gons.
For $n$ odd, they are on ${n-3\over 2}$ concentric $(n-1)$-gons 
while the origin is ${n-1\over 2}$ times degenerate (the latter degeneracy
is lifted by the $y$ coordinate). In either case, the dual generators
reproduce these features. 
Finally, another case that may be discussed easily is the $t'_2$
perturbation by $y$ (see Table II). 

Note again that, like in the $SU(2)$ case, the values taken by 
the perturbed potential in the complex plane can be all linked in a 
connected graph with only straight segments of the same length,
corresponding to the mass of the unique minimal soliton interpolating between
the nearest neighbouring vacua.

Beyond the $SU(3)$ case, it might be possible to investigate some 
(relevant enough) perturbations in the general $SU(N)$, based on 
the following natural conjecture. 
Rearranging the $SU(2)$ type-$A$ perturbed potentials of \DVV, one
can derive the following generating function
\eqn\gensuII{ \sum_{m \geq 0} u^m W_m^{(2)}(x,\Gt_2,\Gt_3,...)=
-\log \big(1-ux+ u^2 \Gt_2+u^3 \Gt_3+...\big) }
where for a given level $k=m-2$, we retain only the couplings
$\Gt_2=t_k$, $\Gt_3=t_{k-1}$,...,$\Gt_{k+2}=t_0$.
The \Che potentials are obtained by taking $\Gt_2=1$, and $\Gt_p=0$
for $p >2$.
Based on the \Che and $(1,0,...,0)$ perturbation cases,
we conjecture that 
this can serve also as a generating function
for certain perturbations of the $SU(N)$ potentials, by 
substituting in \gensuII\ $x \equiv x_1$, $\Gt_2 \equiv x_2$,
$\Gt_3\equiv -x_3$, ...
$\Gt_{N-1} \equiv (-1)^{N-1}x_{N-1}$, and identifying the remaining
couplings as perturbations ($\Gt_p=t_{k+N-p,0,...,0}$ couples to the 
${\cal U}_{k+N-p,0,...,0}$ operator).

\subsec{Fake non ADE cases}
\noindent
Let us return to the few $SU(N)$ cases which avoid 
the appearance of modules. 

\noindent{\it $SU(N)_1$ case.}  It is easy to see that the general
perturbed potential takes the form
$$ W_{N+1}^{(N)}(x_1,...,x_{N-1})=w_{N+1}^{(N)}(x_1,...,x_{N-1})
-\sum_{j=1}^{N-1} s_j x_j -s_0.$$
Working out the perturbed ring, we find that it is isomorphic to the 
perturbed ring of the $SU(2)_{N-1}$ $A_{N}$ theory for some 
special coordinates $s_j$, the identification of the basis
elements being $x_j\equiv(0,..,0,1,0,...,0) \to (j)\equiv x^j$ 
(if the $1$ is in $j^{\rm th}$ position in the $SU(N)_1$ weight). 
\medskip
\noindent{\it $SU(3)_2$ case.}  As mentioned in the previous
section, the perturbed potential, of degree $5$, involves
only couplings of positive dimensions $1,2,3,3,4,5$,
to operators with respective dimensions $4,3,2,2,1,0$. 
(see appendix C for the complete expression.).
The latter match exactly those of the $D_6$ model of $SU(2)$
at level $8$. It is actually straightforward 
to find the isomorphism between the corresponding perturbed rings.
It involves a change of basis of the ring, preserving the 
initial grading (hence allowing for rotations in the two dimensional
space of dimension $2$ ring elements); accordingly the parameters 
$t_{0,0}$, $t_{1,0}$, $t_{1,1}$, $t_{0,2}$ are 
proportional to the $D_6$ parameters $t_0$, $t_2$, 
$t_6$, $t_8$, whereas $t_{2,0}$, $t_{0,1}$ are proportional to
of $t_4\pm i\tau_4$. 
In this sense, $SU(3)_2$ is within 
the $A,D,E$ classification. 

It is then an easy matter to examine what are the normalizable
perturbations of that case, involving only one non-zero flat
coordinate {\it in the $SU(3)$ language}. One finds that the 
solutions are $t_{1,0}\ne 0$, or  $t_{0,1}\ne 0$, or $t_{2,0}\ne 0$
or $t_{0,2}\ne 0$, thus excluding $t_{1,1}$. The first two 
are just particular cases of the discussion above, the 
$t_{2,0}$ perturbation is the ``\Che'' one leading to the $SU(3)_2$
fusion potential, and $t_{0,2}$ is the least relevant perturbation. 
The first and the last have been already found in the discussion of $D_6$.
The perturbation by $t_{2,0}$ that gives a chiral ring isomorphic to
the fusion ring of $SU(3)_2$ is also known to be integrable \CVtat. 
Only the $t_{0,1}$ perturbation had not previously been 
recognized as integrable. In a recent calculation to first order, it has been 
checked that this perturbation admits indeed a spin 3 conserved quantity 
\NWZ.

\newsec{Discussion}
\noindent 
In this paper, we have explored a special class of perturbations of
$N=2$ superconformal theories, in which the basis of the chiral ring is 
made of what we called normalizable matrices. We showed that this 
property is fairly restrictive, and that for the $ADE$ potentials, it 
allows only a finite and small number of perturbations, {\it if we
insist on perturbations in which only one flat coordinate is non-vanishing}. 
We have then shown that this normalizability property leads naturally
to the consideration of the algebra dual to the original chiral one; 
except in a few cases, this dual algebra is well defined and admits a
basis made of matrices with non negative entries (type I) or in which 
at least one matrix has this property (non type I). In all those cases, such
a matrix may be regarded as the adjacency matrix of a graph. The surprising
empirical fact, that generalizes an observation by Lerche and Warner, is 
that this graph resembles the pattern of the extrema of the potential in
coordinate space. This implies that there is a natural action of the
dual algebra on these extrema, namely on the ground states of the theory. 
Finally, a last empirical observation is that there seems to be a connection
between the integrability of the theory and this normalizability condition: 
more precisely, {\it all} the known integrable perturbations of $N=2$
theories with an $ADE$ Landau-Ginzburg potential and a few others have been 
found among the normalizable perturbations. It is tempting 
to conjecture that there is an identity between the two classes. 
In other words, 
normalizability could be a criterion of integrability. 

We now want to discuss this and related questions raised by the 
previous findings. 

\item{*} What is the meaning of the normalizability condition? 
This condition has been introduced on a technical ground, 
namely to allow the diagonalization of the chiral ring in an 
orthogonal basis and the construction of the dual ring. 
Clearly a more physical interpretation would be desirable. 
Let us point out  that this condition is {\it stronger} than the condition
that the 
singularity 
has been fully resolved (in a physical language, that all the 
degeneracy of the extrema has been lifted, and the theory describes 
only massive excitations). Indeed, the normalizability condition
implies  that the coordinates of the extrema are expressible in terms of the 
eigenvectors according to \IIai{}; the independence of the latter
implies the non degeneracy of the former. 
Conversely, it is easy to see that the perturbation by $t_4=t\ne 0$ 
of the $A_6$ potential, {\it v.i.z.} $W={x^7\over 7}-t x^4+t^2 x$
is a full resolution of the $x^7$ singularity, 
but by the theorem of sec. 4.1, 
the matrix $C_1$ of \IVa\ is not normalizable. 

\item{*} What is the good justification of keeping only a single $t\ne 0$?
Certainly the introduction of a single non--zero parameter $t_i$, hence
of a single perturbing operator, is the simplest and most natural thing 
to do. The situation is confused, however, by the existence of some
normalizable cases involving {\it several} non vanishing parameters $\tt$,
presumably non--integrable. 
For example, all the matrices of the chiral ring of the perturbed $A_3$ 
theory
are normalizable for arbitrary $t_1$ and $t_2$. It is very unlikely that all
these perturbations are integrable! 
Yet another example is provided by a class of fusion potentials. 
Whenever a potential is known to be the fusion potential of a rational
conformal field theory, it certainly
satisfies the normalizability condition in a suitable basis. 
Such is the case of the $A_6$ 
potential perturbed by $t_4=1$ and $t_1=2$. In \DFZf, it has been 
showed that this potential $W={x^7\over 7}-x^4-x$ 
provides a one-variable representation of the fusion ring $SU(3)_2$.  
Although the matrix $C_1$ of \IVa\ that encodes the multiplication by $x$ in 
the ordinary basis $1, x, x^2,x^3-1,x^4-2x, x^5-3x^2$ of the $A_6$ case 
is {\it not} normalizable, after a change of basis to the basis 
$$1,x,\oh(x^5-3x^2),\oh( 5x^2-x^5),\oh(x^3-1),\oh(x^4-3x)$$
it reads
\eqn\IVb{C_1=\pmatrix{0 & 1 & 0 & 0 & 0 & 0 \cr
	  	      0 & 0 & 1 & 1 & 0 & 0 \cr
		      1 & 0 & 0 & 0 & 1 & 0 \cr
      		      0 & 0 & 0 & 0 & 1 & 0 \cr
		      0 & 1 & 0 & 0 & 0 & 1 \cr
		      0 & 0 & 1 & 0 & 0 & 0 \cr		   }}
and is normal. Clearly then the $M$ and $N$ algebras are isomorphic (we are 
in a fusion case, hence self-dual). 
It is amusing to see again that the location of extrema in 
the $x$ plane reproduces the pattern of the integrable weights of $SU(3)_2$.
However it is doubtful that this corresponds to an 
integrable perturbation of the $A_6$ theory.
Note that this instance illustrates the possibility of attaching several
consistent gradings to the chiral rings of a given potential. 

Another example is provided by the $Sp(2)_2$ case.  The potentials
for the $Sp(N)_k$ fusion algebra have been worked out \refs{\Bou,\GSc}.
We choose to present now the case of $N=k=2$, due to its relation to 
the $SU(3)_2$ case.
The potential reads
$$W={x^5 \over 5}-x^3y+xy^2 +xy -x.$$
Comparing with the $SU(3)_2$ general perturbations of Appendix C,
we find that this is a special perturbation by ${\cal U}_{1,1}=xy+...$
{\it and} ${\cal U}_{1,0}$ {\it simultaneously}, corresponding to
$t_{1,1}=-1$ and $t_{1,0}=1/2$, the other $t$'s being zero.
As the fusion ring of a WZW model, this point is normalizable, but
corresponds again to a perturbation mixing two directions. In this case
too, the integrability of the $N=2$ theory described by this potential
has not been established, to the best of our knowledge. 

\item{*}
It seems therefore that a refined version of our conjecture should 
be: the normalizability of a perturbation by a single flat coordinate
is equivalent to integrability.

\item{*} Although we have used the language of potentials and 
polynomial representation, it must be clear that the issue of 
normalizability depends only on the structure constants and may therefore
also be addressed in cases where no potential is available. We hope to
return to such instances in a near future.

\item{*} What may be the origin of such an alleged  connection between 
integrability and normalizability? The form of the $C_1$ matrix 
in the simplest cases (see \IVa) suggests a possible connection
with generalized Toda theories and/or hamiltonian reduction. This too
will be left to future investigation.

\item{*} What is  the physical meaning of the graph and/or of the 
dual algebra? The existence of the dual ring, with a basis labelled 
in the same way as the ground states, means that one may define a ring 
structure on these ground states. 
What is the meaning of  this  ring ? 
The whole discussion has some features reminiscent of a recent discussion
by Cecotti and Vafa \CV. These authors have been able to relate the 
counting of solitons (weighted by their fermionic number) interpolating
between pairs of ground states with the intersection numbers of
homology cycles of the (perturbed) potential. 
Their discussion, contrary to ours, is 
not limited to the integrable or normalizable perturbations. With this
additional assumption, we are able to obtain quite explicit 
formulae and new results on the pattern of ground states. It would be
quite interesting to understand if our results have any bearing on that
more general and systematic approach.

\item{*} 
Is there a conformal field theory associated with the integrable cases, 
in the sense that there is a subring of the 
chiral ring isomorphic to the fusion ring of that conformal theory
as in the \Che cases ? 
In all the other $ADE$ cases that we have encountered, there was always a
cyclic $\IZ_N$ subring. In that sense, one may say that there was an
underlying $SU(N)_1$ conformal theory, but it is not clear what is gained 
from that.

\vskip1cm

{\bf Acknowledgements} Special thanks are due to Nick Warner for his 
indefatigable patience in explaining us the beauties and intricacies of
$N=2$ theories. Also, we have benefited from discussions
with D. Gepner, C. Itzykson, W. Lerche, D. Nemeschansky and N. Sochen.
F. Lesage is supported by a Canadian NSERC 67 scholarship.

\nref\Kr{I. Krichever, Comm. Math. Phys. {\bf 143} 415-429 (1992). }
\listrefs

\appendix{A}{The $D_{n+2}$ potential and free energy}
\noindent
The heuristic idea to connect the $D_{n+2}$ and the $A_{2n+1}$ is to
take an orbifold of the latter. We follow here a route slightly 
different from \DVV. Suppose that only the $t_{{\rm even}}$
parameters are non vanishing in the $A_{2n+1}$ potential $W_A(x',\tt)$, 
which is thus an even function of $x'$, $W_A(x',\tt)= V(x'^2,\tt)$. 
We imagine that this potential is used as an action, 
i.e. in an exponentiated form as a weight  in integrals
\eqn\Aa{ \bra f \ket = \int dx' e^{-W_A(x',\tt)} f(x')\ .}
If one restricts oneself to even functions of $x'$, $f(x')=F(x'^2)$, 
one may perform the
$x'^2\to x$ change of variables, and up to irrelevant factors and 
discarding all problems of convergence \dots
\eqnn\Ab
$$\eqalignno{ \bra f \ket &= \int {dx\over \sqrt{x}} e^{-V(x)} F(x) \cr
&= \int dx dy\, e^{-V(x) -\oh y^2 x} F(x) \ .& \Ab\cr }$$
The orbifold $D_{n+2}$ potential is thus identified as the term in the 
exponential; the jacobian of the transformation has forced us 
to introduce 
a new variable $y$, and after inserting an 
extra deformation parameter $\tau$ coupled to $y$, the $D_{n+2}$ potential
reads
\eqnn\Aaa
$$\eqalignno{W_{D_{n+2}}(x,y;t_0,t_2,\cdots,t_{2n},\tau)
&= W_{A_{2n+1}}(x'=\sqrt{x}, 
t_0,0,t_2,\cdots,t_{2n}) + \oh x y^2 +\tau y\ . &\ \cr
&= {x^{n+1}\over 2(n+1)} + \cdots  &\Aaa \cr}$$
It is a quasihomogeneous polynomial of $x$ (degree 2), $y$ (degree $n$) 
and the $t$'s.
The free energy $F$ of these $D$ models may also be determined 
explicitly in terms of the $A$ one. Expressing the multiplication of
the polynomials $p_i(x,\tt)$ and $p'_n=y$ 
in terms of that of the $p^{(A)}_{2i}(x',\tt)$
modulo $\partial_x W,\partial_y W$, one finds that 
\eqn\Ac{
 F_{D_{n+2}}(t_0,t_2,\cdots, t_{2n},\tau)=F_{A_{2n+1}}(t_0,0,t_2,\cdots, t_{2n})
-{\tau^2\over 2} \Phi(t_0,\cdots,t_{2n})\ .}
We recall that expressions for $F_{A}$ or its partial derivatives with 
respect to the $t$'s have been given in \DVV\ and \Kr; 
the function $\Phi$ is a polynomial in $t_0, \cdots, t_{2n}$
that is determined by its partial derivatives: 
$-\partial /\partial t_i \Phi$ is the coefficient of $p_{n-i}$ in the 
expansion of $2V'(x)$. Using the relation $W_{A_{2n+1}}'(x')=p_{2n+1}(x')$
and recursion formulae between the $p$'s, one finds
\eqn\Ad{-\partial /\partial t_{i} \Phi = \sum_{r\ge 0, j_p \ge 0
\atop r+\sum j_p=i}(-1)^r \prod_{p=1}^r t_{2n-2j_p}\ .}

As a side remark, we want to comment the positivity properties
of the coefficients of the resulting $F$. 
While all the monomials of $F_{A}$ have 
positive coefficients, at first sight the monomials of the 
polynomial $\Phi$ seem to have either sign. We have checked in the cases
$D_4$ and $D_6$, and it is very likely to be true for general $D_{n+2}$, 
$n$ {\bf even}, that one may rewrite $F$ with positive signs only in terms
of $t_0, t_2,\cdots, t_{n-2},t_{n+2}, \cdots, t_{2n}$, and  $t_{\pm}=
{1\over \sqrt{2}}(t_{n}\pm i^{{n\over 2}-1}\tau)$. 
This is of interest in view of our 
earlier observation that the chiral ring is a generalization of the
$M$ algebra associated with the graph, and that the latter has non negative 
structure constants only for $D_{{\rm even}}$. It tells 
us what is the change of basis to be performed to deal with an 
algebra with positive structure constants. In contrast for the
$D_{{\rm odd}}$ cases, the signs in $F$ are irreducible. 
One can see that 
the same positivity properties of the coefficients of $F$ holds for
the $A_n$
, $E_6$ and $E_8$ cases \KTS, 
but not for $E_7$. Thus this is one more
manifestation of the type I--non-type I distinction alluded to above.


\appendix{B}{The perturbed potentials of the $E_{6,7,8}$ cases}

\noindent The $E_6$ potential reads
\eqnn\Ba
$$\eqalignno{W&= 
{x^3\over 3}+{y^4\over 4} 
-t_{10}\,x\,y^{2}- t_{7}\,x\,y
-\left(t_{6}-{{t^{3}_{10}}\over{2}}\right)\,y^{2}\cr
& -\left({{t^{4}_{10} }\over{12}}-t_{6}\,t_{10}+t_{4}\right)\,x
-\left(t_{3}-t_{7}\,t^{2}_{10}\right)\,y \cr
& -{{t_{6}\,t^{3}_{10}}\over{6 }}
+{{t_{4}\,t^{2}_{10}}\over{2}}+{{t^{2}_{7}\,t_{10}}\over{2}}+{{t^{2}
 _{6}}\over{2}}-t_{0}\ . & \Ba \cr } $$
\medskip
\noindent The $E_7$ potential is
\eqnn\Bb
$$\eqalignno{W&= 
{x^3\over 3}+x\,y^{3}
-t_{16}\,x^{2}\,y
-\left(t_{12}-{{4\,t^{3}_{16}}\over{27}} \right)\,x^{2}
-\left(t_{10}+{{t^{4}_{16}}\over{27}}-{{4\,t_{12}\,t_{16}}\over{3}}\right)\,x\,y  \cr & \qquad
-t_{8}\,y^{2}
-\left(t_{6}-{{t^{6}_{16}}\over{729}}+{{5\,t_{12}\,t^{3}_{16}}\over{27}}
-{{5\,t_{10}\,t^{2}_{16}}\over{18}}-{{t_{8}\,t_{16}}\over{3}}
-{{5\,t^{2}_{12}}\over{6}}\right)\,x \cr & \qquad
-\left(t_{4}-{{t_{12}\,t^{4}_{16}}\over{162}}
+{{t_{10}\,t^{3}_{16}}\over{54}}+{{t^{2}_{12}\,t_{16}}\over{6}}
-{{t_{6}\,t_{16}}\over{3}}-{{t_{10}\,t_{12}}\over{3}}\right)\,y 
\cr & 
\qquad 
-t_{0}+{{t^{9}_{16}}\over{118098}}-{{t_{12}\,t^{6}_{16}}\over{1458}}
+{{t_{8}\,t^{4}_{16}}\over{81}}+{{t^{2}_{12}\,t^{3}_{16}}\over{27}}
-{{t_{10}\,t_{12}\,t^{2}_{16}}\over{9}}+{{t_{4}\,t^{2}_{16}}\over{9}}
\cr & \qquad 
-{{4\,t_{8}\,t_{12}\,t_{16}}\over{9}}+{{t^{2}_{10}\,t_{16}}\over{9}}
-{{4\,t^{3}_{12}}\over{27}}+{{2\,t_{6}\,t_{12}}\over{3}}
+{{t_{8}\,t_{10}}\over{3}}\ . & \Bb \cr} $$
\medskip
\noindent Finally, the $E_8$ potential reads
\eqn\Bc{W(x,y,\tt)= {x^3\over 3}+{y^5\over 5}-s_{28} xy^3-s_{22}xy^2-s_{18}y^3
-s_{16}xy-s_{12}y^2-s_{10}x-s_{6}y-s_0 }
where
\eqnn\Bd
$$\eqalignno{
s_0&= t_0-t_{28}^{3}\,t_6-{{103\,t_{28}^{15}}\over{450}}
+{{421\,t_{22}\,t_{28}^{11}}\over{450}}
+{{14\,t_{18}\,t_{28}^{9}}\over{45}}-{{29\,t_{16}\,t_{28}^{8}}\over{30}}
-{{43\, t_{23}^{2}\,t_{28}^{7}}\over{45}}
\cr &
+{{7\,t_{12}\,t_{28}^{6}}\over{15}}
+{{4\,t_{10}\,t_{28}^{5}}\over{5}}+{{5\,t_{16}\, t_{22}\,t_{28}^{4}}\over{3}}
+{{t_{22}^{3}\,t_{28}^{3}}\over{3}}
+{{t_{18}^{2}\,t_{28}^{3}}\over{2}}+{{t_{16}\,t_{18}\,t_{28}^{2}}\over{2}}
\cr &
+{{t_{18}\,t_{22}^{2}\,t_{28}}\over{2}}
-t_{10}\,t_{22}\,t_{28}
-{{t_{16}^{2}\,t_{28}}\over{2}}
-{{43\,t_{18}\,t_{22}^{6}}\over{30}}-{{t_{16}\,t_{22}^{2}}\over{2}}
-t_{12}\,t_{18} \cr
s_6&= t_6+{{82\,t_{28}^{12}}\over{75}}-{{107\,t_{22}\,t_{28}^{8}}\over{30}}
-{{7\,t_{18}\,t_{28}^{6}}\over{30}}+{{12\,t_{16}\,t_{28}^{5}}\over{5}}
+{{7\,t_{22}^{2}\,t_{28}^{4}}\over{3}}
-{{4\,t_{12}\,t_{28}^{3}}\over{3}}
\cr & 
\qquad\qquad\qquad +2\,t_{18}\, t_{22}\,t_{28}^{2}
-t_{10}\,t_{28}^{2}-2\,t_{16}\,
 t_{22}\,t_{28}-{{t_{22}^{3}}\over{3}}-t_{18}^{2} \cr
s_{10}&=t_{10} -{{11\,t_{28}^{10}}\over{45}}-{{2\,t_{22}\,t_{28}^{6}
 }\over{15}}+{{3\,t_{18}\,t_{28}^{4}}\over{2}}-{{t_{16}\,
 t_{28}^{3}}\over{2}}+{{t_{22}^{2}\,t_{28}^{2}}\over{2}}-
 t_{12}\,t_{28}-t_{18}\,t_{22}\cr
s_{12}&=t_{12} -{{28\,t_{28}^{9}}\over{15}}+{{23\,t_{22}\,t_{28}^{5}
 }\over{5}}-t_{18}\,t_{28}^{3}-{{3\,t_{16}\,t_{28}^{2}
 }\over{2}}-{{3\,t_{22}^{2}\,t_{28}}\over{2}}
\cr 
s_{16}&=t_{16}+ {{19\,t_{28}^{7}}\over{15}}-{{2\,t_{22}\,t_{28}^{3}
 }\over{3}}-2\,t_{18}\,t_{28}\cr
s_{18}&=t_{18}+ {{6\,t_{28}^{6}}\over{5}}-2\,t_{22}\,t_{28}^{2}\cr
s_{22}&= t_{22}-2\,t_{28}^{4} \cr
s_{28}&= t_{28}\ .& \Bd \cr}$$


\appendix{C}{The $SU(3)_2$ potential and free energy. }

\noindent The $SU(3)_2$ perturbed potential is parametrized as follows

$$\eqalign{ W(x_1,x_2,t_.)&=[{{x_1^{5}}\over{5}}-x_1^{3}\,x_2+x_1\,x_2^{2}]  \cr
 & -{t_{02}}\,x_2^{2}
-{t_{11}}\,x_1\,x_2
-\left( {t_{11}}\,{t_{02}}+{t_{20}}\right)\,x_1^{2}
-\left({t_{01}}-t_{20}- {t_{02}}^{3}-{t_{11}}\,{t_{02}}\right)\,x_2 \cr
& -\left(-{{{t_{11}}\,{t_{02} }^{2}}\over{2}}
+{t_{20}}\,{t_{02}}+{{{t_{11}}^{2}}\over{2}}+{t_{10}}
 \right)\,x_1 \cr
& +{t_{02}}^{2}\,{t_{01}}-{{3\,{t_{02}}^{5}}\over{10}}
 -{t_{20}}\,{t_{11}}-{t_{00}}\ .\cr } $$
The polynomials $p_i(x_1,x_2,t_.)=-\partial W/\partial t_i$ 
form a basis of the chiral ring with structure constants derived from the 
free energy
$$ \eqalign{F&= 
+{{{t_{00}}^{2}\,{t_{02}}}\over{2}}
+{t_{00}}\, {t_{10}}\,{t_{11}}+{{{t_{00}}\,({t_{20}}^{2}+{t_{01}}^{2 })}\over{2}}
+{{{t_{10}}^{2} \,({t_{20}} +{t_{01} })}\over{2}}
-{{{t_{10}}^{2}\,{t_{11}}\,{t_{02}} }\over{2}}
\cr&
 +{{{t_{10}}^{2}\, {t_{02}}^{3}}\over{6}}
-{t_{10}}\,t_{20}\,{t_{01}}\,{t_{02}}
-{{{t_{10}}\,({t_{20}}+t_{01})\,{t_{11}}^{2}}\over{2}}
+{{{t_{10}}\,({t_{20}}+t_{01})\,{t_{11}}\,{t_{02}}^{2}}\over{2}}
\cr &
+{{{t_{10}}\,{t_{11}}^{3}\,{t_{02}}}\over{6}} 
+{{({t_{20}^3}+t_{01}^{3})\,{t_{02}}^{2}}\over{6}}
-{{ {t_{20}}\,t_{01}\,(t_{20}+t_{01})\,{t_{11}}}\over{2}}
+{{({t_{20}}^{2}+t_{20}\,t_{01}+t_{01}^{2})\,{t_{11}}^{2}\,{t_{02}}}\over{2}}
\cr &
+{{({t_{20}}^{2}+t_{01}^{2}) \,{t_{02}}^{5}}\over{20}}
-{{ {t_{20}}\,t_{01}\,{t_{11}}\,{t_{02}}^{3}}\over{2}} 
+{{({t_{20}}+t_{01})\,{t_{11}}^{ 4}}\over{8}}
-{{({t_{20}}+t_{01})\,{t_{11}}^{3}\,{t_{02}}^{2}}\over{ 4}} 
\cr &
+{{({t_{20}}+t_{01})\,{t_{11}}^{2}\,{t_{02}}^{4} }\over{8}} 
%
%
-{{3\,{t_{11}}^{5}\,{t_{02}}}\over{40}}
+{{{t_{11}}^{4}\,{t_{02}}^{3}}\over{8}}
-{{{t_{11}}^{3}\,{t_{02}}^{5}}\over{24}}
+{{{t_{11}}^{2}\,{t_{02}}^{7 }}\over{56}}
+{{{t_{02}}^{11}}\over{3960}}
\cr }$$
and the only non vanishing $\eta_{ij}=\eta_{ji}$ are given by
$$\eqalign{\eta_{(00)(02)}&=\eta_{(10)(11)}=1\cr
\eta_{(20)(20)}&=\eta_{(01)(01)}=1\ .\cr}$$
Note that $F$ is symmetric under the interchange 
$t_{20} \leftrightarrow t_{01}$. 

Upon restriction to $t_{00}=t_{10}=t_{11}=t_{02}=0$, $t_{20}=-1$, $t_{01}=0$, 
the potential reduces to
$$ W={x_1^5\over 5}-x_1^3 x_2+x_1x_2^2+x_1^2-x_2$$
and the polynomials to
$$\Eqalign{p_{00}&=1 & p_{10}&=x_1 \cr p_{20}&= x_1^2-x_2 & 
p_{01}&= x_2 \cr p_{11}&= x_1x_2-1 & p_{02}&= x_2^2-x_1\ ,\cr}$$
that are the polynomials that represent the fusion ring of $SU(3)_2$. 

There is a simple change of variables that maps these expressions 
to those pertaining to $D_6$. Let
$$\Eqalign{x_1&=a(x-t_4)    & x_2&= y+{a^2\over 2}(x^2-3xt_4+3t_4^2-t_3)\cr
t_{02}&=-at_4   & t_{10} &=-a^2t_3 \cr
t_{20}&=-{a^3\over 2}t_2 -a\tau    & t_{01} &= -{a^3\over 2}t_2 +a\tau    \cr
t_{11}&=2 t_1 & t_{02} &=2at_0  \cr }$$
with $a^4=-4$. 
Then $W_{SU(3)_2}(x_1,x_2,t_{02},\dots,t_{00})=W_{D_6}(x,y,t_4,\cdots,t_0)$.

%
\def\pry{y }
\message{The free energies of the $E_6$ and $E_7$ cases have already
appeared in the literature. For completeness and an easier reading, do you
want them printed here together with that of $E_8$ ? (y/n)?}\read-1 to \anspr
\ifx\anspr\pry\message{ALEA JACTA EST 
Make sure that your printer is supplied with sufficient paper !! }
\vfill \eject
\noindent 
1. {\bf The $E_6$ free energy}

\noindent The free energy reads
$$\eqalignno{F&=
 {{t^{13}_{10}}\over{185328}}
+{{t^{2}_{7}\,t^{8}_{10}}\over{576}}
+ {{t^{2}_{6}\,t^{7}_{10}}\over{252}}
+{{t_{6}\,t^{2}_{7}\,t^{5}_{10} }\over{24}}
+{{t^{2}_{4}\,t^{5}_{10}}\over{60}}
+{{t_{4}\,t^{2}_{7}\,t ^{4}_{10}}\over{24}} \cr &
+{{t^{2}_{3}\,t^{4}_{10}}\over{24}}
+{{t^{4}_{7} \,t^{3}_{10}}\over{24}}
+{{t_{3}\,t_{6}\,t_{7}\,t^{3}_{10}}\over{6}}
+ {{t_{4}\,t^{2}_{6}\,t^{3}_{10}}\over{6}}
+{{t^{2}_{6}\,t^{2}_{7}\,t^{ 2}_{10}}\over{4}}
+{{t_{3}\,t_{4}\,t_{7}\,t^{2}_{10}}\over{2}} \cr &
+{{t_{3 }\,t^{3}_{7}\,t_{10}}\over{6}}
+{{t_{4}\,t_{6}\,t^{2}_{7}\,t_{10} }\over{2}}
+{{t^{4}_{6}\,t_{10}}\over{12}}
+{{t^{2}_{3}\,t_{6}\,t_{10} }\over{2}}
+{{t^{3}_{4}\,t_{10}}\over{6}}
+{{t^{2}_{0}\,t_{10}}\over{2 }} \cr &
+{{t_{6}\,t^{4}_{7}}\over{12}}
+{{t^{2}_{4}\,t^{2}_{7}}\over{4}}
+{{ t_{3}\,t^{2}_{6}\,t_{7}}\over{2}}
+t_{0}\,t_{3}\,t_{7}+t_{0}\,t_{4}\, t_{6}
+{{t^{2}_{3}\,t_{4}}\over{2}}\ .\cr } $$
\bigskip
\noindent {2. {\bf The $E_7$ free energy}}

$$ \eqalignno{F &=
{{t^{19}_{16}}\over{1001094543576}}
+{{t^{2}_{12}\,t^{13}_{16}}\over{ 55269864}}
+{{t^{2}_{10}\,t^{11}_{16}}\over{5196312}} \cr &
-{{t^{3}_{12}\,t^{10}_{16}}\over{1417176}}
+{{t_{10}\,t^{2}_{12}\,t^{9}_{16}}\over{157464}}
+{{t^{2}_{8}\,t^{9}_{16}}\over{236196}}
-{{t_{8}\,t^{2}_{12}\,t^{8}_{16}}\over{52488}} \cr &
+{{t^{2}_{10}\,t_{12}\,t^{8}_{16}}\over{104976}}
+{{7\,t^{4}_{12}\,t^{7}_{16}}\over{157464}}
+{{t_{8}\,t_{10}\,t_{12}\,t^{7}_{16} }\over{8748}}
+{{t^{3}_{10}\,t^{7}_{16}}\over{52488}}\cr &
+{{t^{2}_{6}\,t^{7}_{16}}\over{10206}}
-{{7\,t_{10}\,t^{3}_{12}\,t^{6}_{16}}\over{52488}}
+ {{t_{6}\,t_{10}\,t_{12}\,t^{6}_{16}}\over{2916}}
-{{t^{2}_{8}\,t_{12}\,t^{6}_{16}}\over{2916}}
-{{t_{8}\,t^{2}_{10}\,t^{6}_{16}}\over{5832}}\cr &
+{{t_{8}\,t^{3}_{12}\,t^{5}_{16}}\over{972}}
+{{t^{2}_{10}\,t^{2}_{12}\,t^{5}_{16}}\over{972}}
-{{t_{6}\,t_{8}\,t_{12}\,t^{5}_{16}}\over{486}}
+{{t_{6}\,t^{2}_{10}\,t^{5}_{16}}\over{972}}
+{{t^{2}_{4}\,t^{5}_{16}}\over{810}}\cr &
-{{7\,t^{5}_{12}\,t^{4}_{16}}\over{11664}}
+{{5\,t_{6}\,t^{3}_{12}\,t^{4}_{16}}\over{2916}}
-{{5\,t_{8}\,t_{10}\,t^{2}_{12}\,t^{4}_{16}}\over{972}}
+{{5\,t^{3}_{10}\,t_{12}\,t^{4}_{16}}\over{5832}}
+{{t_{4}\,t_{8}\,t_{12}\,t^{4}_{16}}\over{162}}\cr &
+{{t^{2}_{6}\,t_{12}\,t^{4}_{16}}\over{324}}
+{{t_{4}\,t^{2}_{10}\,t^{4}_{16}}\over{648}}
+{{t_{6}\,t_{8}\,t_{10}\,t^{4}_{16}}\over{162}}
-{{t^{3}_{8}\,t^{4}_{16}}\over{486}}
+{{25\,t_{10}\,t^{4}_{12}\,t^{3}_{16}}\over{5832}}\cr &
-{{t_{4}\,t^{3}_{12}\,t^{3}_{16}}\over{486}}
+{{t^{2}_{8}\,t^{2}_{12}\,t^{3}_{16}}\over{54}}
+{{t_{8}\,t^{2}_{10}\,t_{12}\,t^{3}_{16}}\over{162}}
+{{t_{4}\,t_{6}\,t_{12}\,t^{3}_{16}}\over{27}}
+{{t^{4}_{10}\,t^{3}_{16}}\over{486}}\cr &
-{{t_{4}\,t_{8}\,t_{10}\,t^{3}_{16}}\over{54}}
+{{t^{2}_{6}\,t_{10}\,t^{3}_{16}}\over{54}}
-{{t_{8}\,t^{4}_{12}\,t^{2}_{16}}\over{72}}
-{{t^{2}_{10}\,t^{3}_{12}\,t^{2}_{16}}\over{324}}
+{{t_{4}\,t_{10}\,t^{2}_{12}\,t^{2}_{16}}\over{36}}\cr &
+{{t_{6}\,t_{8}\,t^{2}_{12}\,t^{2}_{16}}\over{18}}
+{{t_{6}\,t^{2}_{10}\,t_{12}\,t^{2}_{16}}\over{18}}
-{{t^{2}_{8}\,t_{10}\,t_{12}\,t^{2}_{16}}\over{18}}
-{{t^{2}_{4}\,t_{12}\,t^{2}_{16}}\over{18}}
+{{t_{4}\,t_{6}\,t_{10}\,t^{2}_{16}}\over{18}}\cr &
+{{t_{4}\,t^{2}_{8}\,t^{2}_{16}}\over{18}}
-{{t^{2}_{6}\,t_{8}\,t^{2}_{16}}\over{18}}
+{{t^{6}_{12}\,t_{16}}\over{486}}
-{{t_{6}\,t^{4}_{12}\,t_{16}}\over{108}}
+{{t_{8}\,t_{10}\,t^{3}_{12}\,t_{16}}\over{18}}\cr &
+{{t^{3}_{10}\,t^{2}_{12}\,t_{16}}\over{54}}
-{{t_{4}\,t_{8}\,t^{2}_{12}\,t_{16}}\over{6}}
+{{t^{2}_{6}\,t^{2}_{12}\,t_{16}}\over{9}}
-{{t_{6}\,t_{8}\,t_{10}\,t_{12}\,t_{16}}\over{9}}
+{{2\,t^{3}_{8}\,t_{12}\,t_{16}}\over{27}}\cr &
+{{t_{6}\,t^{3}_{10}\,t_{16}}\over{27}}
+{{t^{2}_{8}\,t^{2}_{10}\,t_{16}}\over{18}}
+{{t^{2}_{4}\,t_{10}\,t_{16}}\over{6}}
+{{t_{4}\,t_{6}\,t_{8}\,t_{16}}\over{3}}
+{{t^{3}_{6}\,t_{16}}\over{9}}\cr &
+{{t^{2}_{0}\,t_{16}}\over{2}}
-{{t_{10}\,t^{5}_{12}}\over{216}}
+{{t_{4}\,t^{4}_{12}}\over{72}}
+{{t_{6}\,t_{10}\,t^{3}_{12}}\over{18}}
-{{2\,t^{2}_{8}\,t^{3}_{12}}\over{27}}\cr &
-{{t_{8}\,t^{2}_{10}\,t^{2}_{12}}\over{18}}
-{{t_{4}\,t_{6}\,t^{2}_{12}}\over{6}}
+{{t^{4}_{10}\,t_{12}}\over{108}}
+{{t_{4}\,t_{8}\,t_{10}\,t_{12}}\over{3}}
+{{t^{2}_{6}\,t_{10}\,t_{12}}\over{6}}
+{{t_{6}\,t^{2}_{8}\,t_{12}}\over{3}}\cr &
+t_{0}\,t_{4}\,t_{12}
+{{t_{4}\,t^{3}_{10}}\over{18}}
-{{t^{3}_{8}\,t_{10}}\over{18}}
+t_{0}\,t_{6}\,t_{10}
-{{t_{0}\,t^{2}_{8}}\over{2}}
-{{t^{2}_{4}\,t_{8}}\over{2}}
+{{t_{4}\,t^{2}_{6}}\over{2}} \cr} $$


%
\noindent 3. {\bf{The $E_8$ free energy}}

\noindent
The function $F$ has the following simple expression
$$ \eqalignno{
F&={{t_{28}^{7}\,t_6^{2}}\over{210}}
+{{t_{22}\,t_{28}^{3}\,t_6^{2}}\over{6}}
+{{t_{18}\,t_{28}\,t_6^{2} }\over{2}}
+{{t_{16}\,t_6^{2}}\over{2}}
+t_0\,\left(  t_{22}\,t_6+t_{10}\,t_{18}+t_{12}\,t_{16}\right) 
\cr &
+ t_{16}\,\left({{t_{18}\,t_{28}^{6}}\over{60}}+{{t_{18}\,
 t_{22}\,t_{28}^{2}}\over{2}}+{{t_{18}^{2}}\over{2}}\right) \,t_6
+{{t_{12}\,t_{22}\,t_{28}^{6}\,t_6}\over{45 }}
+{{t_{18}\,t_{22}^{2}\,t_{28}^{5}\,t_6}\over{60}}
\cr &
+{{ t_{10}\,t_{22}\,t_{28}^{5}\,t_6}\over{15}}
+{t_{16}^{2}\over{2}}\,\left({{t_{28}^{5}}\over{20}}+t_{22}\,t_{28}\right)
\,t_6
+t_{16}\,
\left({{t_{22}^{2}\,t_{28}^{4} }\over{12}}+{{t_{22}^{3}}\over{6}}\right)\,t_6
\cr & 
+{{t_{12}\, t_{18}\,t_{28}^{4}\,t_6}\over{6}}
+{{t_{22}^{4}\,  t_{28}^{3}\,t_6}\over{72}}
+{{t_{18}^{2}\,t_{22}\,t_{28} ^{3}\,t_6}\over{6}}
+{{t_{12}\,t_{16}\,t_{28}^{3}\, t_6}\over{6}}
+{{t_{12}\,t_{22}^{2}\,t_{28}^{2}\,  t_6}\over{2}}
\cr 
&
+{{t_{10}\,t_{16}\,t_{28}^{2}\,t_6}\over{ 2}}
+{{t_{18}\,t_{22}^{3}\,t_{28}\,t_6}\over{6}}
+{{ t_{10}\,t_{22}^{2}\,t_{28}\,t_6}\over{2}}
+{{t_{12} ^{2}\,t_{28}\,t_6}\over{2}}
+t_{12}\,t_{18}\,t_{22} \,t_6
\cr &
+t_{10}\,t_{12}\,t_6
+{{t_{28}^{31}}\over{ 245764125000}}
+{{t_{22}^{2}\,t_{28}^{23}}\over{27945000}}
\cr &
+ t_{22}^{3}\,\left({{t_{28}^{19}}\over{729000}}
+{{2\,t_{22} \,t_{28}^{15}}\over{30375}}
+{{13\,t_{22}^{2}\,t_{28}^{11} }\over{10800}}
+{{11\,t_{22}^{3}\,t_{28}^{7}}\over{1080}}
+{{ t_{22}^{4}\,t_{28}^{3}}\over{54}}\right)
\cr &
+t_{18}^{2}\,\left( {{t_{28}^{19}}\over{1539000}}
+{{t_{22}^{2}\,t_{28}^{11} }\over{360}}
+{{7\,t_{22}^{3}\,t_{28}^{7}}\over{180}}
+{{17\, t_{22}^{4}\,t_{28}^{3}}\over{72}}\right)
+{{t_{16}^{2}\, t_{28}^{17}}\over{459000}}
\cr &
+t_{18}\,t_{22}^{2}\,\left({{ t_{28}^{17}}\over{81000}}
+{{7\,t_{22}\,t_{28}^{13}}\over{ 16200}}
+{{77\,t_{22}^{2}\,t_{28}^{9}}\over{6480}}
+{{7\,t_{22}^{3}\,t_{28}^{5}}\over{90}}
+{{t_{22}^{4}\,t_{28}}\over{ 18}}\right)
\cr &
+{{t_{16}\,t_{22}^{2}\over{2}}\,\left({{t_{28}^{16} }\over{40500}}
+{{13\,t_{22}\,t_{28}^{12}}\over{8100}}
+{{ t_{22}^{2}\,t_{28}^{8}}\over{30}}
+{{7\,t_{22}^{3}\,t_{28}^{4} }\over{36}}
+{{2\,t_{22}^{4}}\over{45}}\right)}
\cr &
+t_{16} \,t_{18}\,t_{22}\,\left({{t_{28}^{14}}\over{5400}}
+{{11\, t_{22}\,t_{28}^{10}}\over{1800}}
+{{7\,t_{22}^{2}\,t_{28} ^{6}}\over{45}}
+{{11\,t_{22}^{3}\,t_{28}^{2}}\over{24}}\right)
\cr &
+{{t_{18}^{3}\,\over{6}}\left({{t_{28}^{13}}\over{1800}}
+{{t_{18}\, t_{28}^{7}}\over{20}}
+{{7\,t_{22}^{2}\,t_{28}^{5}}\over{10 }}
+t_{22}^{3}\,t_{28}
+{{3\,t_{18}^{2}\,t_{28}}\over{10}} \right)}
\cr &
+{{t_{16}^{2}\,t_{22}\,\over{2}}\left({{t_{28}^{13} }\over{5400}}
+{{t_{22}\,t_{28}^{9}}\over{60}}
+{{7\,t_{22}^{ 2}\,t_{28}^{5}}\over{30}}
+{{5\,t_{22}^{3}\,t_{28}}\over{12 }}\right)}
\cr &
+{{t_{12}^{2}\,\over{2}}\left({{t_{28}^{13}}\over{ 8775}}
+{{t_{22}\,t_{28}^{9}}\over{270}}
+{{t_{22}^{2}\,t_{28}^{5}}\over{5}}
+{{2\,t_{22}^{3}\,t_{28}}\over{3}}\right) }
\cr &
+t_{12}\,t_{18}\,t_{22}\,\left({{t_{28}^{12} }\over{2700}}
+{{t_{22}\,t_{28}^{8}}\over{60}}
+{{5\,t_{22}^{ 2}\,t_{28}^{4}}\over{18}}
+{{t_{22}^{3}}\over{6}}\right)
\cr &
+t_{12}\,t_{16}\,t_{22}\,\left({{t_{28}^{11}}\over{900}}
+{{2\,t_{22}\,t_{28}^{7}}\over{45}}
+{{4\,t_{22}^{2}\,t_{28}^{3}}\over{9}}\right)
\cr &
+{{t_{16}^{2}\,t_{18}\,\over{2}}\left({{t_{28}^{11}}\over{1800}}
+{{t_{22}\,t_{28}^{7}}\over{15}}
+{{3\,t_{18} \,t_{28}^{5}}\over{20}}
+{{7\,t_{22}^{2}\,t_{28}^{3}}\over{6 }}
+t_{18}\,t_{22}\,t_{28}\right)}
\cr &
+t_{10}^{2}\, \left({{t_{28}^{11}}\over{4950}}
+{{t_{18}\,t_{28}^{5} }\over{20}}
+{{t_{22}^{2}\,t_{28}^{3}}\over{6}}
+{{t_{10}\, t_{28}}\over{6}}
+{{t_{16}\,t_{22}}\over{2}}\right)
\cr &
+t_{10}\,t_{16}\,\left({{t_{22}\,t_{28}^{10}}\over{900}}
+{{7\, t_{22}^{2}\,t_{28}^{6}}\over{180}}
+{{t_{22}^{3}\,t_{28} ^{2}}\over{3}}\right)
\cr &
+t_{12}\,t_{22}^{3}\,\left({{11\,t_{29 }^{10}}\over{8100}}
+{{7\,t_{22}\,t_{28}^{6}}\over{270}}
+{{ t_{22}^{2}\,t_{28}^{2}}\over{12}}\right)
\cr 
&
+t_{16}^{3}\,\left( {{t_{28}^{10}}\over{2700}}
+{{7\,t_{22}\,t_{28}^{6}}\over{ 360}}
+{{t_{18}\,t_{28}^{4}}\over{24}}
+{{t_{16}\,t_{28}^{ 3}}\over{24}}
+{{t_{22}^{2}\,t_{28}^{2}}\over{4}}
+{{t_{18}\, t_{22}}\over{3}}\right)
\cr &
+t_{12}\,t_{16}\,t_{18}\,\left({{ t_{28}^{9}}\over{180}}
+{{t_{22}\,t_{28}^{5}}\over{6}}
+{{ t_{18}\,t_{28}^{3}}\over{3}}
+{{3\,t_{22}^{2}\,t_{28} }\over{2}}\right)
\cr &
+t_{10}\,t_{18}^{2}\,\left({{t_{28}^{9} }\over{360}}
+{{t_{18}\,t_{28}^{3}}\over{6}}
+{{t_{22}^{2}\, t_{28}}\over{2}}\right)
+t_{10}\,t_{22}^{3}\,\left({{t_{28}^{9}}\over{1620}}
+{{7\,t_{22}\,t_{28}^{5}}\over{360}}
+{{ t_{22}^{2}\,t_{28}}\over{30}}\right)
\cr &
+{{t_{16}\,t_{18}^{2 }\over{2}}\,\left({{t_{22}\,t_{28}^{8}}\over{20}}
+{{2\,t_{22}^{2}\, t_{28}^{4}}\over{3}}
+t_{18}\,t_{22}\,t_{28}^{2}
+{{2\, t_{22}^{3}}\over{3}}\right)}
\cr &
+t_{10}\,t_{12}\,t_{22}\,\left({{t_{28}^{8}}\over{90}}
+{{t_{22}\,t_{28}^{4} }\over{6}}
+{{t_{22}^{2}}\over{3}}\right)
+t_{10}\,t_{18}\, \left({{t_{22}^{2}\,t_{28}^{7}}\over{45}}
+{{t_{22}^{3}\, t_{28}^{3}}\over{6}}\right)
\cr &
+{{t_{12}^{2}\,t_{18}\,\over{2}}\left({{ t_{28}^{7}}\over{90}}
+{{2\,t_{22}\,t_{28}^{3}}\over{3}}
+ t_{18}\,t_{28}\right)}
+t_{10}\,t_{16}^{2}\, \left({{t_{28}^{7}}\over{120}}
+{{t_{22}\,t_{28}^{3}}\over{6 }}
+{{t_{18}\,t_{28}}\over{2}}
+{{t_{16}}\over{6}}\right)
\cr &
+{{ t_{12}\,t_{18}^{2}\over{2}}\,\left({{7\,t_{22}\,t_{28}^{6}}\over{ 45}}
+t_{22}^{2}\,t_{28}^{2}
+{{2\,t_{18}\,t_{22}}\over{3 }}\right)}
+t_{12}^{2}\,t_{16}\,\left({{t_{28}^{6} }\over{45}}
+{{t_{22}\,t_{28}^{2}}\over{2}}
+{{t_{18}}\over{2 }}\right)
\cr &
+t_{10}\,t_{12}\,\left(t_{16}\,\left({{t_{28}^{ 5}}\over{30}}
+t_{22}\,t_{28}\right)
+{{t_{12}\,t_{28}^{3} }\over{6}}
+t_{18}\,t_{22}\,t_{28}^{2}\right)
+t_{10}\, t_{16}\,\left({{t_{18}\,t_{22}\,t_{28}^{4}}\over{3}}
+{{ t_{18}\,t_{22}^{2}}\over{2}}\right)
\cr &
+t_{12}\,t_{16}^{2}\, \left({{t_{22}\,t_{28}^{4}}\over{4}}
+{{t_{18}\,t_{28}^{2 }}\over{2}}
+{{t_{16}\,t_{28}}\over{6}}
+{{t_{22}^{2}}\over{2 }}\right)
+{{t_0^{2}\,t_{28}}\over{2}}+{{t_{12}^{3}\,t_{22}}\over{3}} \ .
\cr}$$
\else\vskip1mm
\fi

%
%
%
%
%
%
\vfill\eject
\input epsf

\centerline{\bf Table captions}

\bigskip
\bigskip
\bigskip

\noindent{\bf Table I: A, D, E perturbations.}

\noindent{}Normalizable perturbations of the $A$, $D$, $E$
potentials are displayed as follows. First column: name
of the potential; second column: name of the perturbation
(list of the non--zero $t$'s); third column: the dual ring
of the normalized ring, through the graph of one of its generators
(all the cycles have to be understood as oriented anti--clockwise);
fourth column: locus of the extrema of the perturbed potential
($x$ in the complex plane in the one variable case, $x$ and $y$
planes otherwise); fifth column: values taken by the perturbed potential
at the various extrema, in the complex plane (the links correspond 
to the minimal solitons interpolating between the extrema);
sixth column: a check--mark in case of known integrability, a question--mark
otherwise. 
 
\bigskip
\bigskip

\noindent{\bf Table II: SU(3) perturbations.}

\noindent{}Normalizable perturbations of the diagonal 
${\cal A}^{(n=k+3)}$ series of the Kazama--Suzuki cosets
${SU(3)_k\times SO(4)_1 \over SU(2)_{k+1} \times U(1)}$. The columns
are organized as on Table I. The first perturbation 
$t_k=t_{n-3}\equiv t_{k,0,...,0}$
is the ``Chebishev" one, the corresponding third column 
displays the weight diagram of $SU(3)_k$ (generalization of the 
$A$ Dynkin diagram) and should be understood as 
oriented in order for each elementary triangle to be itself 
oriented anti--clockwise. The other cycles of the third column have to be 
understood as oriented anti--clockwise.

\vfill\eject
{\par\begingroup\parindent=0pt\leftskip=1cm
\rightskip=1cm\parindent=0pt
\baselineskip=11pt
\epsfxsize=12cm
\midinsert
\centerline{\epsfbox{f1.eps}}
\bigskip
\centerline{\bf Table I.}
\par
\endinsert
\endgroup
\par}
\vfill\eject
{\par\begingroup\parindent=0pt\leftskip=1cm
\rightskip=1cm\parindent=0pt
\baselineskip=11pt
\epsfxsize=12cm
\midinsert
\centerline{\epsfbox{f2.eps}}
\bigskip
\centerline{\bf Table I (continued).}
\par
\endinsert
\endgroup
\par}
\vfill\eject
{\par\begingroup\parindent=0pt\leftskip=1cm
\rightskip=1cm\parindent=0pt
\baselineskip=11pt
\epsfxsize=12cm
\midinsert
\centerline{\epsfbox{f3.eps}}
\bigskip
\centerline{\bf Table I (continued).}
\par
\endinsert
\endgroup
\par}
\vfill\eject
{\par\begingroup\parindent=0pt\leftskip=1cm
\rightskip=1cm\parindent=0pt
\baselineskip=11pt
\epsfxsize=12cm
\midinsert
\centerline{\epsfbox{f4.eps}}
\bigskip
\centerline{\bf Table II.}
\par
\endinsert
\endgroup
\par}

\bye